\documentclass[review]{elsarticle}

\usepackage[version=3]{mhchem}
\usepackage{amsmath}
\usepackage{multirow}
\usepackage{subcaption}
\usepackage{hyperref}

\usepackage{cleveref}

\usepackage{array}
\makeatletter
\newcommand{\thickhline}{%
	\noalign {\ifnum 0=`}\fi \hrule height 1pt
	\futurelet \reserved@a \@xhline}
\newcolumntype{"}{@{\hskip\tabcolsep\vrule width 1pt\hskip\tabcolsep}}
\makeatother 

\biboptions{numbers,sort&compress}

\usepackage{color} 

\crefname{figure}{figure}{figures}
\crefname{equation}{equation}{equations}
\crefdefaultlabelformat{(#2#1#3)} 

\begin{document}

\begin{frontmatter}


\title{Non-dimensional analysis of electrochemical governing equations of lead-acid batteries}



\author[mymainaddress]{Tayyeb Nazghelichi}
\ead{tnazghelichi@mail.kntu.ac.ir}

\author[mymainaddress]{Farschad Torabi\corref{mycorrespondingauthor}}
\cortext[mycorrespondingauthor]{Corresponding author}
\ead{ftorabi@kntu.ac.ir}

\author[mysecondaryaddress]{Vahid Esfahanian}
\ead{evahid@ut.ac.ir}

\address[mymainaddress]{Faculty of Mechanical Engineering, K. N. Toosi University of Technology, Tehran, Iran}
\address[mysecondaryaddress]{Vehicle, Fuel and Environment Research Institute, School of Mechanical Engineering, College of Engineering, University of Tehran, Tehran, Iran
}

\begin{abstract}
Lead--acid batteries have widespread usages in various kinds of industries all over the world. Therefore, these electrochemical energy saving devices always need to develop new models and analyses based on innovative ideas. In the present study, a new set of non--dimensional electrochemical governing equations for lead--acid cells are derived based on an innovative self--comparative concept. Thorough this non--dimensionalization process, some useful dimensionless coefficients of the governing equations are introduced. The non--dimensionalization analysis has been applied to the electrochemical governing equations  including conservation of charge in solid and electrolyte, and conservation of species.
Four novel dimensionless coefficients of electrode conductivity, electrolyte conductivity, diffusional conductivity of species and diffusion coefficient are derived from the dimensionless model. The new dimensionless model is studied in analysis of two distinct case. Firstly, in comparison of dimensional and non--dimensional models for two typical lead--acid cells. Secondly, in analysis of a single cell with a set of experimental test. Both cases simulated using finite volume method. The simulated data is validated using comparison with experimental data. Finally, shown results indicate that the non--dimensional model is in fairly good accordance with data obtained from experiments, moreover, dimensionless coefficients are useful for comparing purposes and analysis of electrochemical processes. In conclusion, this study demonstrates that investigation of lead--acid cell's performance in comparison with it's maximum potential is appropriate and operational method.

\end{abstract}


\end{frontmatter}


\section{Introduction}
The world's environmental and economical future is predicted to be influenced by production or consumption of energy, related to limited resources of fossil fuels. Electrochemical energy due to independency of fossil fuels, zero--emission of air pollutants and sustainability is under attention nowadays. Batteries supply energy of electrical devices on demand via storage and conversion of chemical energy. Estimation of battery market shows domination of lead-acid batteries in the rechargeable market~\cite{bkbtry}. Lead--acid batteries have many advantages comparing other rechargeable batteries such as working on higher voltages, worthy specific energy that is energy per unit mass, operation over a wide temperature range that means lower need to thermal management, low cost of maintenance and manufacturing and having one of the most successful recycling systems on the world~\cite{zhuetal11}. 

Utilization of lead--acid batteries covers a wide variety of obligations for different roles, from high current quick pulse to lower sustained current, from internal combustion engine to backup power of telecommunications. Also, deep discharges and recharges over short periods of time in electric vehicles are supposed to be tolerated. Thus, the battery is expected to provide enough power for the defined functions as well~\cite{vrlabk}. 

In all the above--mentioned cases, obtaining a proper model for analysis of battery behavior under a wide variety of different tasks is one of the main themes of studies. Modeling and simulation are a way to analyze the problem numerically and could bring a better perception of physics of events~\cite{esbabak}. Many researchers have  been interested in modeling and simulation of lead-acid batteries and the reviewed studies can be divided into three major divisions as follows.

  The first division is about fundamental procedures for modeling of phenomena and processes of battery functions and deriving governing equations in whole cell or different parts of it. Newman and Tiedemann~\cite{newmantiedman75} reviewed   developments of simulation  in primary  and  secondary  batteries in regard to  the  theory  of  flooded  porous  electrodes. Further, they developed equations  to  provide  a  basis  for  examining  the  behavior  of specific systems  such  as  primary  and  secondary  batteries. Gu et al.~\cite{guetal87} developed a model to study the state of charge~(SoC) of a cell during discharge, rest and charge. The model was capable to predict the dynamic behavior of acid concentration  and porosity of electrodes. Nguyen et al.~\cite{nguyenetal90} used a volume--averaging technique and concentrated binary electrolyte theory to model the transport of electrolyte and investigated the effect of separator design on the discharge of starved lead--acid cells. Vidts and White~\cite{vidtswhite97} derived general governing equations that can be used to model mass transport and ohmic drop in porous electrodes containing three phases of solid, liquid and gas. A micro--macroscopic coupled model  accounts for the effects of microscale and interfacial non--equilibrium processes on the macroscopic species and charge transfer was developed by Wang et al.~\cite{wangetal98}. Moreover, Catherino at al.~\cite{catherinoetal99} worked on a general method to model the curve of constant current charging. They showed that the gas evolution process occurring at a constant voltage is independent of the normally occurring gas evolution process on the electrode surface at higher voltages and appears as a kinetically controlled gas evolution step.   
  
Further studies on fundamental modeling of batteries have been conducted as follows. Torabi and Esfahanian~\cite{torabiesfahanian11} investigated thermal--runaway (TRA) as one of the battery failure modes. They developed a general set of governing equations by which the thermal behavior of batteries could be obtained. The presented approach could be used for investigating the thermal--runaway in any kind of battery systems. In another study~\cite{torabiesfahanian13}, they detail the main sources of heat generation in lead--acid batteries. They introduced a new phrase called general Joule heating, considering similarity between irreversible heat and Joule effect.
Oury et al.~\cite{ouryetal14} proposed a stationary model to predict the electrochemical behavior of a cell in which  honeycomb--shaped positive \ce{PbO2}--electrode  were sandwiched between two planar negative electrodes.  Their results showed that the positive current distribution is nearly completely specified by effects of geometry, with little influence from the hydrodynamic.  Recently, Zhung et al.~\cite{zhungetal17} introduced an advanced methodology for modeling of battery state estimation. The  conventional techniques calibrated the parameters of state estimation during development stage before vehicle production, while, different usage of a battery result in different aging processes. Literature review showed that some other works could be placed in the first division too~\cite{kimhong00,guwang00,tennoetal01,wangsrinivasan02,guetal02,bernhard03,srinivasanetal03,gandhi15,nandanwarkumar16,pilatowiczetal16}.

The second division of studies is the usage of various known mathematical methods for modeling and improving the models time cost as well as precision. Ball et al.~\cite{balletal02} used finite element method for modeling the current density of the valve regulated lead--acid battery on the positive grid. Esfahanian and Torabi~\cite{esfahaniantorabi06} applied Keller--Box method for simulation of transport equations in lead--acid batteries numerically. They indicated that the Keller--Box method is a suitable method for integration of electrochemical transport equations both in integrated  and  multi--region  formulation. Shen~\cite{shen07} utilized neural network to  the battery residual available capacity estimation in terms of the  state of available capacity for electric vehicles. He approved effectiveness of the  state of available capacity by comparison of experimental data and proposed neural network. An improved model based on computational
fluid dynamics (CFD) and equivalent circuit model was introduced by Esfahanian et al.~\cite{esfahanianetal08}. They reported the approach is very fast and accurate. Vasebi et al.~\cite{vasebietal08} developed a novel model  based on the extended Kalman filter  for estimating the SoC. Moreover, Burgos et al.~\cite{burgosetal15} used fuzzy modeling for the SoC estimation. They indicated that the performance of the model is better than that obtained from conventional models. Esfahanian et al.~\cite{esfahanianetal15} investigated a reduced order model based on the proper orthogonal decomposition method  to the coupled one--dimensional electrochemical transport equations. Furthermore, Ansari et al.~\cite{ansarietal16} applied the similar technique in order to reduce the computational time suitable for real--time monitoring  purposes.

The third division can be dedicated for applications of lead--acid batteries  in renewable and hybrid energy systems such as photovoltaic and wind power systems  as well as  automotive. Albers~\cite{albers09} investigated grid corrosion and water loss as main effects of high heat into starter batteries. The investigation showed that AGM batteries perform much better than flooded batteries under high temperature condition. Fendri and Chaabene~\cite{fenderichaabene12} developed a dynamic model for estimating the open circuit voltage to follow the SoC of a lead--acid battery connected to photovoltaic panel. Zhang et al.~\cite{zhangetal11} developed a new model to investigate dynamics of lead--acid batteries for automotive applications. Moreover, they proposed an integrated method for battery state of health monitoring. A coulomb counting method was developed to evaluate SoC of a gelled lead--acid battery by Gonzalez et al.~\cite{gonzalezetal12} to control a hybrid system of wind--solar test--bed with hydrogen support. Dufo-Lopez et al.~\cite{lopezetal14} investigated  components of a photovoltaic system specially battery charge controller. They used a weighted Ah--throughput method to provide more accurate lifetime values. Silva and Hendrick~\cite{silvahendrick16} analyzed self--sufficiency of household lead--acid battery coupled with photovoltaic system and its possible interaction with the grid. There are some researches in this section could be find by literature review~\cite{ achaibouetal12, blaifietal16,yuanetal15,titoetal16,dufolopezetal16}.

In the all reviewed literature, governing equations of lead--acid batteries have been investigated dimensionally and very little information is available on the non--dimensional analysis~\cite{alavyoonetal91,srinivasanetal03,mcguinness03}. In the mentioned references, the non--dimensionalization applied to equations like Navier--Stokes and some unique parameters such as acid concentration but electrochemical governing equations have been used dimensionally. However, the non--dimensional analyze of electrochemical equations has important advantages that has been investigated in this study.

In fact, the importance and advantages of non--dimensionalization of electrochemical governing equations of lead--acid batteries were neglected in previous researches. Firstly, dimensionless variations and coefficients are needed in some features such as comparisons between different batteries and in control systems. Secondly, non--dimensionalization could improve the accuracy and stability of simulations because of reducing errors and simulation time through normalized scaling, instead of working with measured parameters. Eventually, advancements in analyses of battery modeling could be achieved as well as experimental results. Moreover, derived non--dimensional numbers are expected to play a major role in some investigations such as instability studies.
The main objective of the present study is  non--dimensionalization of electrochemical equations governing on lead--acid batteries and introduce some new proper dimensionless numbers. Furthermore, simulation of the system using CFD method and comparison of obtained results are additional purposes. In the present study, some new definitions in batteries investigation have been developed that necessarily not limited to only lead--acid ones.

\section{Mathematical formulation}
As mentioned in the previous section, Wang et al.~\cite{wangetal98} developed the general micro--macro model of battery dynamics. In the present study the electrolyte assumed to be immobilized using gelled electrolyte. So, the following equations applied to the non--dimensionalization  process. Equation~\eqref{phiseq} shows conservation of charge in solid:	
\begin{equation}
\nabla.(\sigma^{\mathrm{eff}} \nabla \phi_s)=Aj 
\label{phiseq}
\end{equation}
The conservation of charge in electrolyte can be displayed as:
\begin{equation}
 \nabla.(k^{\mathrm{eff}}\nabla \phi_e)+\nabla.(k_D^{\mathrm{eff}}\nabla \ln{c})=-Aj 
 \label{phileq}
\end{equation}
and the following equation shows conservation of species:
\begin{equation}
 \varepsilon \frac{\partial c}{\partial t} = \nabla.(D^{\mathrm{eff}}\nabla c)+ a_2\frac{A j}{2F}  
 \label{ceq}
\end{equation}
 The term $j$ is the transfer current density and can be calculated from the general Butler--Volmer relation:
\begin{equation}
j= i_0 \left( \frac{c}{c_{\mathrm{ref}}} \right)^{\gamma} \left\{ \exp \left( \frac{\alpha_a F}{RT}\eta \right) - \exp \left(- \frac{\alpha_c F}{RT}\eta \right) \right\} 
\label{butlervolmer}
\end{equation}

All the parameters are defined in the list of symbols in nomenclature section. Non--dimensionalization is the removal of units from an equation including physical quantities through suitable substitution of variables. The open circuit at fully charged state (OCFCS) is suggested as proper state for non--dimensionalization of governing equations by the authors. The OCFCS is an equilibrium state, containing maximum level of energy and applied as an appropriate criterion for comparison of battery states during discharging (or charging) process. So, the obtained dimensionless terms refer to intrinsic quantities of the system. The proper parameters were suggested as below (the asterisk sign ($\ast$) shows dimensionless variables):
\begin{itemize}
\item dimensionless electric potential of solid and electrolyte:
\begin{equation}
\phi_s^{\ast}=\frac{\phi_s}{V_{\mathrm{oc,0}}} \Rightarrow \phi_s=V_{\mathrm{oc,0}}\phi_s^{\ast}
\label{phisdless}
\end{equation}
\begin{equation}
\phi_e^{\ast}=\frac{\phi_e}{V_{\mathrm{oc,0}}} \Rightarrow \phi_e=V_{\mathrm{oc,0}} \phi_e^{\ast}
\label{phildless}
\end{equation}

\item dimensionless electrolyte concentration:
\begin{equation}
c^{\ast}=\frac{c}{c_0} \Rightarrow c=c_0c^{\ast}
\label{cdless}
\end{equation}
\item dimensionless cell--length:
\begin{equation}
x^{\ast}=\frac{x}{L} \Rightarrow x=Lx^{\ast} 
\label{xdless}
\end{equation}
\item dimensionless transfer current density:
\begin{equation}
j^{\ast}=\frac{j}{i_0} \Rightarrow j=i_0j^{\ast}
\label{jdless}
\end{equation}
\item dimensionless form of activated area:
\begin{equation}
A^{\ast}=\frac{A}{A_{\text{max}}} \Rightarrow A=A_{\text{max}}A^{\ast} 
\label{adless}
\end{equation}
\item dimensionless time can be defined as: 
\begin{equation}
t^{\ast}=\frac{t}{\tau} \Rightarrow t=\tau t^{\ast}
\end{equation}
in which:
\begin{equation}
\tau=\frac{Fc_0}{i_0 A_{\mathrm{max}}}
\label{tau}
\end{equation}

The variable $\tau$, is a key parameter in non--dimensionalization of electrochemical governing equations that can be used to calculate different relative times for different batteries. The dimensionless time, $t^{\ast}$, can be obtained from other ways but \cref{tau} derived as the proper one in the present investigation. The variable $\tau$ could be called as charge transfer time (CTT) and could be defined as needed time to transfer all existing charge in a unit volume with rate of $i_0$. By this definition, CTT is different for any distinct battery, resulting in different $t^{\ast}$ for them.

\end{itemize}
 By replacement the set of \cref{phisdless,phildless,cdless,xdless,jdless,adless}  into equation~\eqref{phiseq}, one can obtain:
\begin{equation}
\nabla.\left(\frac{V_{\mathrm{oc,0}} \sigma^{\mathrm{eff}}}{i_0A_{\mathrm{max}}L^2}\nabla \phi_s^{\ast}\right)=A^{\ast}j^{\ast}
\label{phis1}
\end{equation}
Thus, dimensionless conductivity of solid yields:
\begin{equation}
\sigma^{\ast}=\frac{V_{\mathrm{oc,0}} \sigma^{\mathrm{eff}}}{i_0A_{\mathrm{max}}L^2}
\label{sigmaast}
\end{equation}
Also, the ~\cref{phis1} can be rewritten as desirable form of:
\begin{equation}
\nabla.(\sigma^{\ast}\nabla \phi_s^{\ast})=A^{\ast}j^{\ast}
\label{phisast}
\end{equation}
Similarly, substituting \cref{phisdless,phildless,cdless,xdless,jdless,adless} into ~\cref{phileq} leads to:
\begin{equation}
\begin{split}
 \nabla.\left(\frac{V_{\mathrm{oc,0}}k^{\mathrm{eff}}}{i_0A_{\mathrm{max}}L^2}\nabla \phi_e^{\ast}\right)+\nabla.\left(\frac{V_{\mathrm{oc,0}}k_\mathrm{D}^{\mathrm{eff}}}{i_0A_{\mathrm{max}}L^2}\nabla \ln{c_0}\right)+& \\
 \nabla.\left(\frac{V_{\mathrm{oc,0}}k_\mathrm{D}^{\mathrm{eff}}}{i_0A_{\mathrm{max}}L^2}\nabla \ln{c^{\ast}}\right)
 &=-A^{\ast}j^{\ast}
 \end{split}
\end{equation}
The second term on the left--hand side expected to be zero because the initial concentration ($c_0$) assumed to be constant over the domain at initial state, therefore:
\begin{equation}
 \nabla.\left(\frac{V_{\mathrm{oc,0}} k^{\text{eff}}}{i_0A_{\text{max}} L^2} \nabla \phi_e^\ast\right)+\nabla.\left(\frac{k_\mathrm{D}^{\text{eff}}}{i_0A_{\text{max}} L^2} \nabla \ln{c^\ast}\right)=-A^\ast j^\ast 
\end{equation}
Now, two more dimensionless numbers of electrolyte conductivity and diffusion of species can be determined:
\begin{equation}
k^{\ast}=\frac{V_{\mathrm{oc,0}}k^{\mathrm{eff}} }{i_0A_{\mathrm{max}} L^2}
\label{kast}
\end{equation}
\begin{equation}
k_\mathrm{D}^{\ast}=\frac{ k_\mathrm{D}^{\text{eff}} }{i_0A_{\mathrm{max}} L^2}
\label{kdast}
\end{equation}
Therefore, final dimensionless form of conservation of charge in electrolyte can be written as:
\begin{equation}
\nabla.(k^{\ast}\nabla \phi_e^\ast)+\nabla.(k_\mathrm{D}^{\ast}\nabla \ln{c^\ast})=-A^\ast j^\ast 
\label{philast}
\end{equation}
Finally, for non--dimensionalization of ~\cref{ceq} the same technique has been applied:
\begin{equation}
 \varepsilon \frac{\partial c^\ast}{\partial t^\ast} = \nabla.(\frac{Fc_0 D^{\mathrm{eff}}}{i_0 A_{\mathrm{max}}  L^2}\nabla c^{\ast})+ \frac{a_2}{2} A^\ast j^\ast 
\end{equation}
Likewise, dimensionless diffusion coefficient obtained:
\begin{equation}
D^{\ast}=\frac{Fc_0 D^{\mathrm{eff}}}{i_0 A_{\mathrm{max}}  L^2}
\label{dast}
\end{equation}
and the unitless form of ~\cref{ceq} became:

\begin{equation}
 \varepsilon \frac{\partial c^\ast}{\partial t^\ast} = \nabla.(D^{\ast}\nabla c^{\ast})+ \frac{a_2}{2} A^\ast j^\ast 
 \label{cast}
\end{equation}

In summary, non--dimensional equations of~\eqref{phisast}, \eqref{philast} and \eqref{cast} with new defined dimensionless parameters of~\eqref{sigmaast}, \eqref{kast}, \eqref{kdast} and \eqref{dast} used for simulations.

\section{Physical interpretation}

Equation~\eqref{sigmaast} can be write down in three eligible forms in order to better perception. This equation can be regarded as: 
\begin{equation}
\sigma^{\ast}=\frac{V_{\mathrm{oc,0}}\sigma^{\mathrm{eff}}\Big/L}{i_0 A_{\mathrm{max}}L}=\frac{i_{\mathrm{oc(s)}}}{i_{\mathrm{exchange}}}
\label{sigmaast1}
\end{equation}
which presents the ratio of conductive current density of solid--electrode to exchange current density (ECD). In fact, the numerator is a hypothetical current density exerting by open circuit voltage. Also, the denominator is the product of ECD and $A_{\mathrm{max}}L$. The latter is a new dimensionless property which can be named ``dimensionless volume''. The dimensionless volume describes the effect of geometrical parameters on $\sigma^{\ast}$.
Obviously, the open circuit voltage of initial time and the effective conductivity of solid are in direct relation with $\sigma^{\ast}$. Conversely, activated area, ECD and cell length are in the inverse relation. The cell length has the most effect on values of $\sigma^{\ast}$.

The non--dimensionalize conductivity of solid can be viewed in another way:

\begin{equation}
\sigma^{\ast}=\frac{V_{\mathrm{oc,0}}}{i_0 A_{\mathrm{max}} L^2 \Big/ \sigma^{\mathrm{eff}}}= \frac{V_{\mathrm{oc,0}}}{V_{\mathrm{exchange(s)}}}
\label{sigmaast2}
\end{equation}
The above fraction is the ratio of $V_{\mathrm{oc},0}$ to exchange voltage of solid. The exchange voltage can be defined as a motive force within solid causing exchanged current under changing $\sigma^{\mathrm{eff}}$ condition. Interestingly, the other form of  \cref{sigmaast} is related to material properties:
\begin{equation}
\sigma^{\ast}= \frac{\sigma^{\mathrm{eff}}}{i_0A_{\mathrm{max}}L^2 \Big/ V_{\mathrm{oc,0}}}=\frac{\sigma^{\mathrm{eff}}}{\sigma_{\mathrm{exchange}}}
\label{sigmaast3}
\end{equation}

Thus, \cref{sigmaast3} is the ratio of the effective conductivity of solid to exchange conductivity, defined in the denominator, and will be discussed more, later in the present paper.

Likewise, \cref{philast} can be rewritten into three suitable forms:

\begin{equation}
k^{\ast}=\frac{V_{\mathrm{oc,0}}k^{\mathrm{eff}} \Big/ L }{i_0 A_{\mathrm{max}} L}=\frac{i_{\mathrm{oc(e)}}}{i_{\mathrm{exchange}}}
\label{kast1}
\end{equation}
Equation~\eqref{kast1} shows the ratio of conductive current density of electrolyte to ECD.
\begin{equation}
k^{\ast}= \frac{V_{\mathrm{oc,0}} }{i_0 A_{\mathrm{max}} L^2 \Big/ k^{\mathrm{eff}}}=\frac{V_{\mathrm{oc,0}}}{V_{\mathrm{exchange(e)}}} 
\label{kast2}
\end{equation}
 Equation~\eqref{kast2} indicates relation of $V_{\mathrm{oc},0}$ to exchange voltage of electrolyte.
\begin{equation}
k^{\ast}= \frac{  k^{\mathrm{eff}}}{i_0 A_{\mathrm{max}} L^2 \Big/V_{\mathrm{oc,0}}}=\frac{k^{\mathrm{eff}}}{k_{\mathrm{exchange}}}
\label{kast3}
\end{equation}
and \cref{kast3} represents proportion of effective conductivity of electrolyte to exchange conductivity.  It is obvious that the exchange conductivity of electrode and electrolyte is equal, considering \cref{kast3,sigmaast3}, and it could be called as exchange conductivity (EC):
\begin{equation}
\mathrm{EC}=\sigma_{\mathrm{exchange}}=k_{\mathrm{exchange}}
\end{equation}
Therefore, \cref{kast3,sigmaast3} can be rewritten as:
\begin{equation}
\sigma^{\ast}=\frac{\sigma^{\mathrm{eff}}}{\mathrm{EC}}
\label{sigmaec}
\end{equation}
\begin{equation}
k^{\ast}=\frac{k^{\mathrm{eff}}}{\mathrm{EC}}
\label{kec}
\end{equation}
Hence, by equating $\mathrm{EC}$ from \cref{sigmaec,kec} the following expression can be calculated as:
\begin{equation}
\mathrm{EC}= \frac{k^{\mathrm{eff}}}{k^{\ast}}= \frac{\sigma^{\mathrm{eff}}}{\sigma^{\ast}}
\label{ksigmaec}
\end{equation}
Also, from \cref{sigmaast2,kast2} one can obtain:
\begin{equation}
\frac{V_{\mathrm{exchange(s)}}}{V_{\mathrm{exchange(e)}}}=\frac{k^{\ast}}{\sigma^{\ast}}
\end{equation}
that is a relation between exchange voltage and conductivity of both electrode and electrolyte. The following equation illustrates the convenience form of  dimensionless diffusional conductivity of species:
\begin{equation}
k_D^{\ast}=\frac{k_D^{\text{eff}}\Big/ L}{i_0 A_{\mathrm{max}}L}
\label{kdast1}
\end{equation}
that shows the ratio of diffusional current density of species to exchange current density. According to \cref{ksigmaec,kdast1}, a notable relationship between effective conductivity of electrode, effective conductivity of electrolyte and diffusional conductivity of species can be obtained using defined dimensionless coefficients:
\begin{equation}
\mathrm{EC}=\frac{\sigma^{\mathrm{eff}}}{\sigma^{\ast}}=\frac{k^{\mathrm{eff}}}{k^{\ast}}=\frac{k_\mathrm{D}^{\mathrm{eff}}}{k_\mathrm{D}^{\ast}V_{\mathrm{oc,0}}}
\label{ec}
\end{equation}

EC and $V_{\mathrm{oc},0}$ are constant numbers for each battery and can be easily calculated from battery characteristics. Equation~\eqref{ec} gives useful relation between coefficients of $\sigma^{\mathrm{eff}}$, $k^{\mathrm{eff}}$, and $k_\mathrm{D}^{\mathrm{eff}}$ by having numerical solutions for $\sigma^{\ast}$, $k^{\ast}$ and $k_\mathrm{D}^{\ast}$.

By applying Similar approach to \cref{dast} the first appropriate form of dimensionless diffusion coefficient can be rewritten as:
\begin{equation}
D^{\ast}=\frac{c_0 D^{\mathrm{eff}}/L^2}{i_0 A_{\mathrm{max}}/F}
\end{equation}
In fact, this equation is a fraction of diffusional molar flow rates. The numerator is in relation to initial concentration and the denominator is related to OCFCS.
The second form is relation of diffusivity:
\begin{equation}
D^{\ast}=\frac{D^{\mathrm{eff}}}{i_0 A_{\mathrm{max}} L^2\Big/ Fc_0}=\frac{D^{\mathrm{eff}}}{D_{\mathrm{exchange}}}
\label{dast1}
\end{equation}

This fraction indicates the ratio of effective diffusion coefficient to exchange diffusion coefficient that could be defined as diffusion coefficient in OCFCS.

Finally, the third form can be represented as:
\begin{equation}
D^{\ast}=\frac{c_0}{i_0 A_{\mathrm{max}}  L^2\Big/ FD^{\mathrm{eff}}}=\frac{c_0}{c_{\mathrm{exchange}}}
\label{dast2}
\end{equation}
The denominator of above fraction could be explained as exchange concentration, that is an abstract concept, resulted from some parameters of OCFCS and dimensionless diffusion coefficient. The parameter of $D^{\ast}$ can be interpreted as a concentration of charge due to changing diffusion rate. 

\begin{figure}
	\centerline{\includegraphics[trim=0cm 7cm 0cm 7cm, clip, width=1\textwidth]{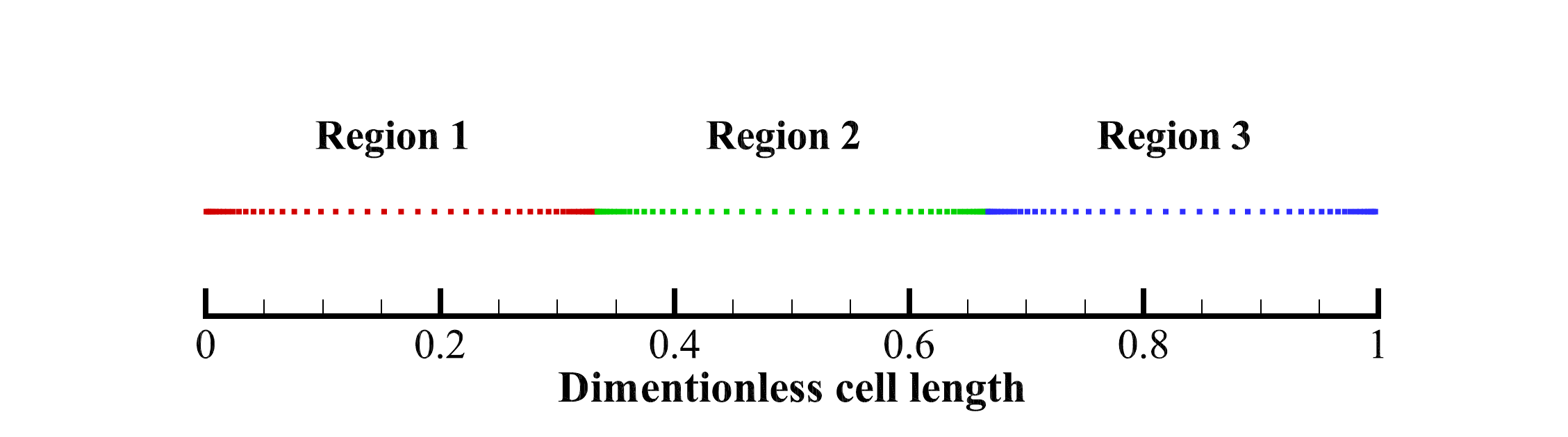}}
	\caption{The meshed domain of solution}
	\label{cell0}
\end{figure}

\begin{table}
	  \caption{Grid independency}
\centering
\begin{tabular}{l*{4}{c}}
\hline
  \thickhline			
  Case & 1 & 2 & 3 &  4  \\
	  \thickhline
  Number of nodes & 15 & 45 & 135 & 275  \\

Dimensionless  concentration, $c^{\ast}$ & 0.725 & 0.705 &0.699 &0.698 \\
  \thickhline  
\end{tabular}
    \label{grid}
\end{table}

In the following sections, two case studies have been conducted based on the newly introduced non--dimensional model, with labels of case study I and case study II. The former one compared non--dimensional simulation results of two different cells, while the later one discussed about discharge processes in one cell experimentally and numerically.

\section{Case studies}
\label{cases}
In this section, the importance of the present study is indicated firstly by solving two sample batteries, illustrated in table~\eqref{samplebatteries}, numerically. Secondly, a set of experimental tests have been conducted to analyze effects of discharge rates on the battery behavior non--dimensionally. In this regard,  one lead--acid battery is discharged at constant currents of $C_1$, $C_2$ and $C_3$, then it is simulated based on properties shown in \cref{propex}. The finite volume method has been applied to simulate the governing equations. The domain of numerical solution assumed to be one--dimensional as the height and width of usual electrode plates are much more than the thickness. The domain consisted of three regions including positive electrode as Region--1, electrolyte as Region--2 and negative electrode as Region--3, which represented in \cref{cell0} in dimensionless scale. A non--uniform mesh generated for each region to optimize accuracy and computational time. A grid independency test was performed by examining electrolyte concentration at the mid point of Region--1. In \cref{grid}, results of grid study is shown. As can be seen, the difference between case 2 and 3 is less than 1\%, so the grid 2 has been selected for all solutions,  saving cost and time ensuring the solution and results be grid independent.

For both cases of I and II, numerical analysis conducted in discharge state of battery in a constant current. Thus, the initial and boundary conditions can be represented as follows:
\begin{itemize}
\item initial conditions of non--dimensional potential in solid and electrolyte obtained by solving whole domain with a very small time step about $10^{-8}$ second.
\item initial condition of dimensionless acid concentration can be calculated using given parameters:
\begin{equation}
c^{\ast}=c_0^{\ast}
\end{equation}
\item boundary conditions of solid:
\begin{equation}
\phi_s^{\ast}=0 \quad  , \quad  x^{\ast}=0
\end{equation}
\begin{equation}
-\sigma^{\ast}\frac{\partial \phi_s^{\ast}}{\partial x^{\ast}}=\mathrm{I} \quad  , \quad  x^{\ast}=1
\end{equation}
\item boundary conditions of liquid:
\begin{equation}
\frac{\partial \phi_e^{\ast}}{\partial x^{\ast}}= 0 \quad , \quad x^{\ast}=0,1
\label{bcphil}
\end{equation}
\item boundary conditions of acid concentration:
\begin{equation}
\frac{\partial c^{\ast}}{\partial x^{\ast}}= 0 \quad , \quad x^{\ast}=0,1
\label{bcc}
\end{equation}
\end{itemize}

 Both \cref{bcphil,bcc} shows the symmetry boundary conditions as points $x^{\ast}=0$ and $x^{\ast}=1$ assumed to be in the center of electrodes.

\begin{table}
  \caption{Input parameters of simulated cells of Case study I}
\centering
\begin{tabular}{ l  c  c c  c  c c }
  			
  Cell & \multicolumn{3}{c}{I (Ref.\cite{guetal87})} & \multicolumn{3}{c}{II (Ref.\cite{alavyoonetal91})}  \\
  \thickhline 
  Initial acid concentration ($c_0$), mol cm$^{-3}$ &\multicolumn{3}{c}{$4.9$e$-3$ } & \multicolumn{3}{c}{$2$e$-4$}  \\

Initial acid concentration ($c_0$), g cm$^{-3}$ &\multicolumn{3}{c}{$0.4806$ } & \multicolumn{3}{c}{$0.0196$}  \\

Initial open circuit voltage ($V_{\mathrm{oc},0}$), volt &\multicolumn{3}{c}{$2.12$ } & \multicolumn{3}{c}{$1.83$}  \\
  
Operating temperature, $^{\circ}$C & \multicolumn{3}{c}{$25$} & \multicolumn{3}{c}{$25$} \\
    
  Transfer number of H$^+$ & \multicolumn{3}{c}{$0.72$} & \multicolumn{3}{c}{$0.80$}  \\
  
  Applied current density ($I_{\text{app}}$), mA cm$^{-2}$ & \multicolumn{3}{c}{$-340$} & \multicolumn{3}{c}{$-9.343$}  \\
  \multirow{2}{*}{Regions } & 1 & 2 & 3 & 1 & 2 & 3 \\
  &(pos)&(sep)&(neg)&(pos)&(sep)&(neg) \\
  \cline{2-7}
  Region width, cm & 0.06 & 0.06 & \multicolumn{1}{c}{0.06} & 0.2&0.2&0.2  \\
  
  Porosity & 0.53 & 0.73 & \multicolumn{1}{c}{0.53} & 0.5& 0.9 & 0.5  \\
  
  Transfer current density ($i_0$), mA cm$^{-2}$ & 10 & - & \multicolumn{1}{c}{10} & 0.1 & - & 0.1  \\
  
   Maximum activated area ($A_{\text{max}}$), cm$^2$cm$^{-3}$ & 100 &  - & \multicolumn{1}{c}{100} & 100 & - &  100 \\
  
   Maximum capacity ($Q_{\text{max}}$), C cm$^{-3}$ & 5660 & - & \multicolumn{1}{c}{5660} & 3130 & - & 3700  \\
  
 Exponent in Butler--Volmer equation ($\gamma$)   & 1.5 & - & \multicolumn{1}{c}{1.5} & 1.5 & - & 1.5 \\
  
   Apparent transfer coefficient for anode ($\alpha_a$) & 0.5 & -  &\multicolumn{1}{c}{0.5} &1 & - &1  \\
  
      Apparent transfer coefficient for cathode ($\alpha_c$) & 0.5 & - & \multicolumn{1}{c}{0.5} & 1& -& 1  \\
  \thickhline
\end{tabular}
    \label{samplebatteries}
\end{table}

\begin{table}
	\caption{Properties used in numerical solution for Case study II} 
	\centering
	\begin{tabular}{ l  c  c c  c  c }
		
		Properties & \multicolumn{3}{c}{Values}   \\
		\thickhline 
		Initial acid concentration ($c_0$), mol cm$^{-3}$ &\multicolumn{3}{c}{$5$e$-3$ } \\
Initial acid concentration ($c_0$), g cm$^{-3}$ &\multicolumn{3}{c}{$0.4903$ } \\
		Initial open circuit voltage ($V_{\mathrm{oc},0}$), volt &\multicolumn{3}{c}{$2.1$ } \\
		Operating temperature, $^{\circ}$C & \multicolumn{3}{c}{$25$} \\
		
		Transfer number of H$^+$ & \multicolumn{3}{c}{$0.72$}  \\
		
		Applied current density ($I_{\text{app}}$), mA cm$^{-2}$ & \multicolumn{3}{c}{$16$ , $24$ , $48$}  \\
		
		 \multirow{2}{*}{Regions } & 1 & 2 & 3  \\
		 &(pos)&(sep)&(neg)\\
		\cline{2-4}
		Region width, cm & 0.098 & 0.0915 & \multicolumn{1}{c}{0.098}   \\
		
		Porosity & 0.6 & 0.8 & \multicolumn{1}{c}{0.6}   \\
		
		Transfer current density ($i_0$), mA cm$^{-2}$ & 4 & - & \multicolumn{1}{c}{4}  \\
		
		Maximum activated area ($A_{\text{max}}$), cm$^2$cm$^{-3}$ & 150 &  - & \multicolumn{1}{c}{150} \\
		
		Maximum capacity ($Q_{\text{max}}$), C cm$^{-3}$ & 7200 & - & \multicolumn{1}{c}{7200} \\
		
		Exponent in Butler--Volmer equation ($\gamma$)   & 1.2 & - & \multicolumn{1}{c}{1.2} \\
		
		Apparent transfer coefficient for anode ($\alpha_a$) & 0.5 & -  &\multicolumn{1}{c}{0.5} \\
		
		Apparent transfer coefficient for cathode ($\alpha_c$) & 0.5 & - & \multicolumn{1}{c}{0.5}  \\
		\thickhline
	\end{tabular}
	\label{propex}
\end{table}

In case study I, performance of non--dimensional model versus dimensional model have been discussed through comparison of two different cell. Moreover, in case study II performance of non--dimensional model have been analyzed for one cell under various discharge currents.

\section{Results and discussion}
\begin{figure}
	\begin{subfigure}{0.5\textwidth}\centerline{\includegraphics[trim= 0mm 0mm 0mm 70mm, width=1.1\textwidth]{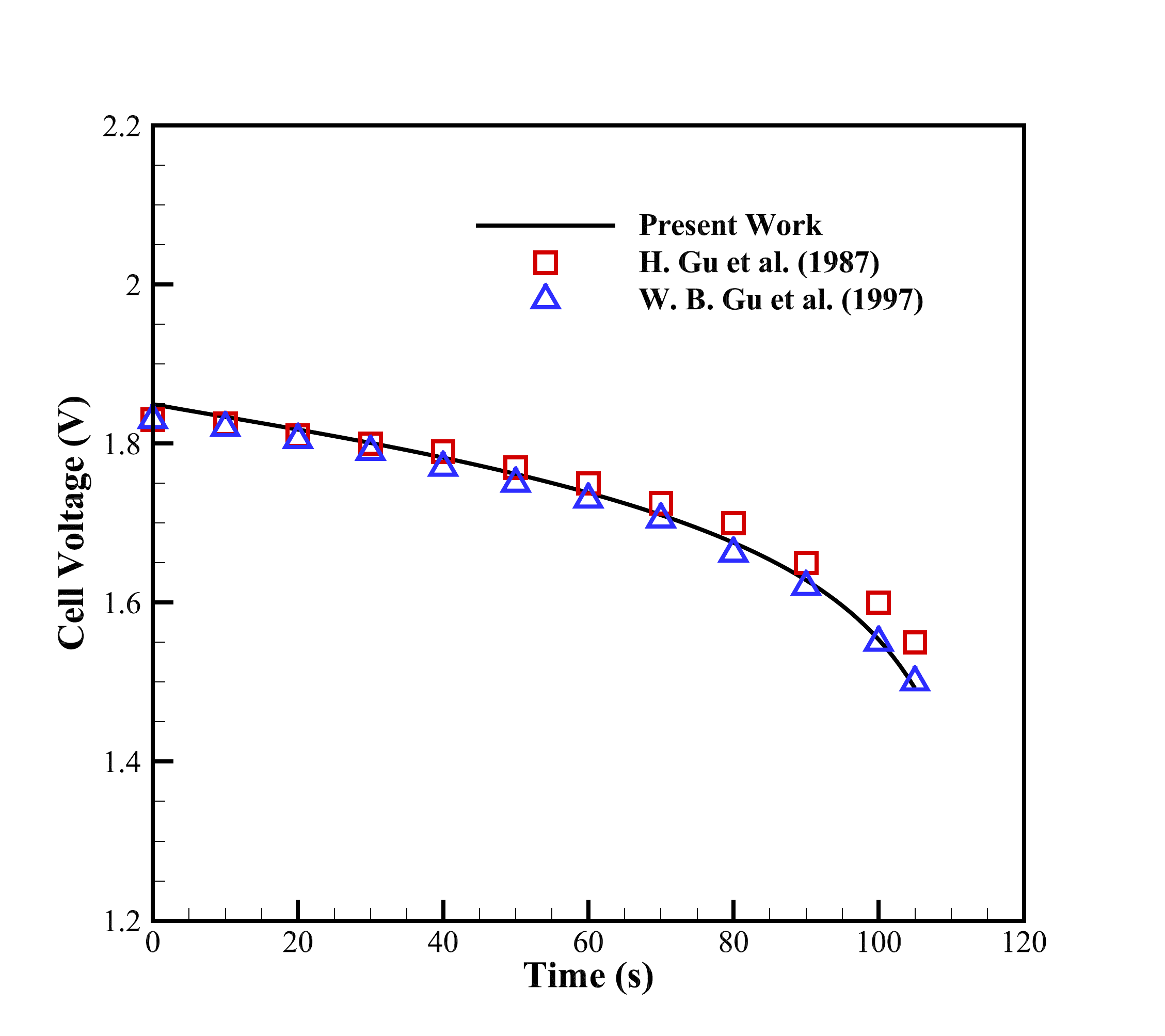}}
		\caption{case I}
		\label{valid1}
	\end{subfigure}
\begin{subfigure}{0.5\textwidth}\centerline{\includegraphics[trim= 0mm 0mm 0mm 70mm, width=1.1\textwidth]{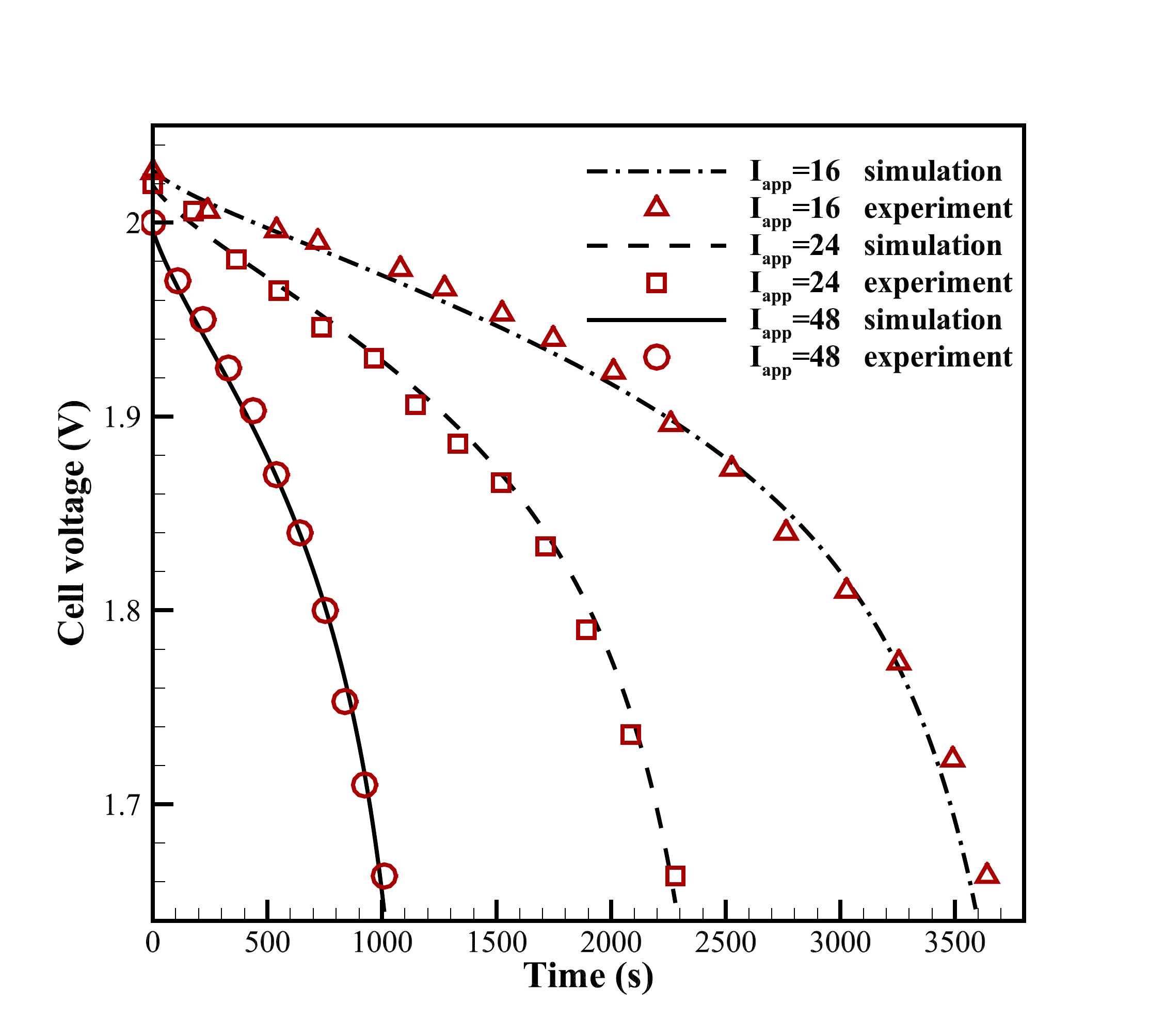}}
	\caption{case II}
	\label{valid2}
\end{subfigure}
\caption{ Validation test: Voltage of cell during discharge. Case (I) comparison of two literature works with present work at a constant current. Case (II) comparison of experimental data with simulated data in three discharge constant currents. }
\label{validation}
\end{figure}

\begin{figure}
\centerline{\includegraphics[trim= 0mm 0mm 0mm 60mm, width=0.5\textwidth]{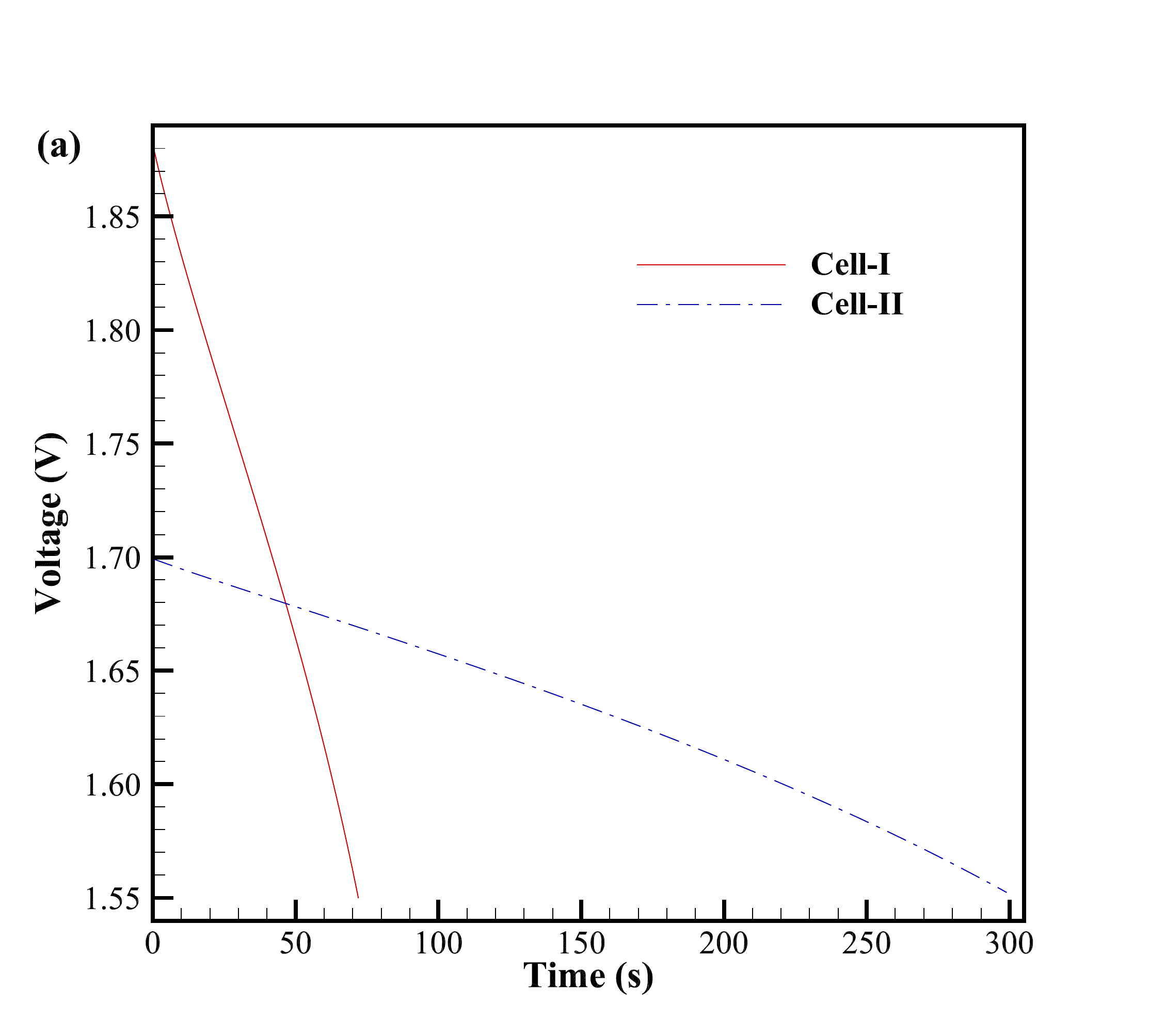}}
\centerline{\includegraphics[trim= 0mm 0mm 0mm 20mm, clip, width=0.5\textwidth]{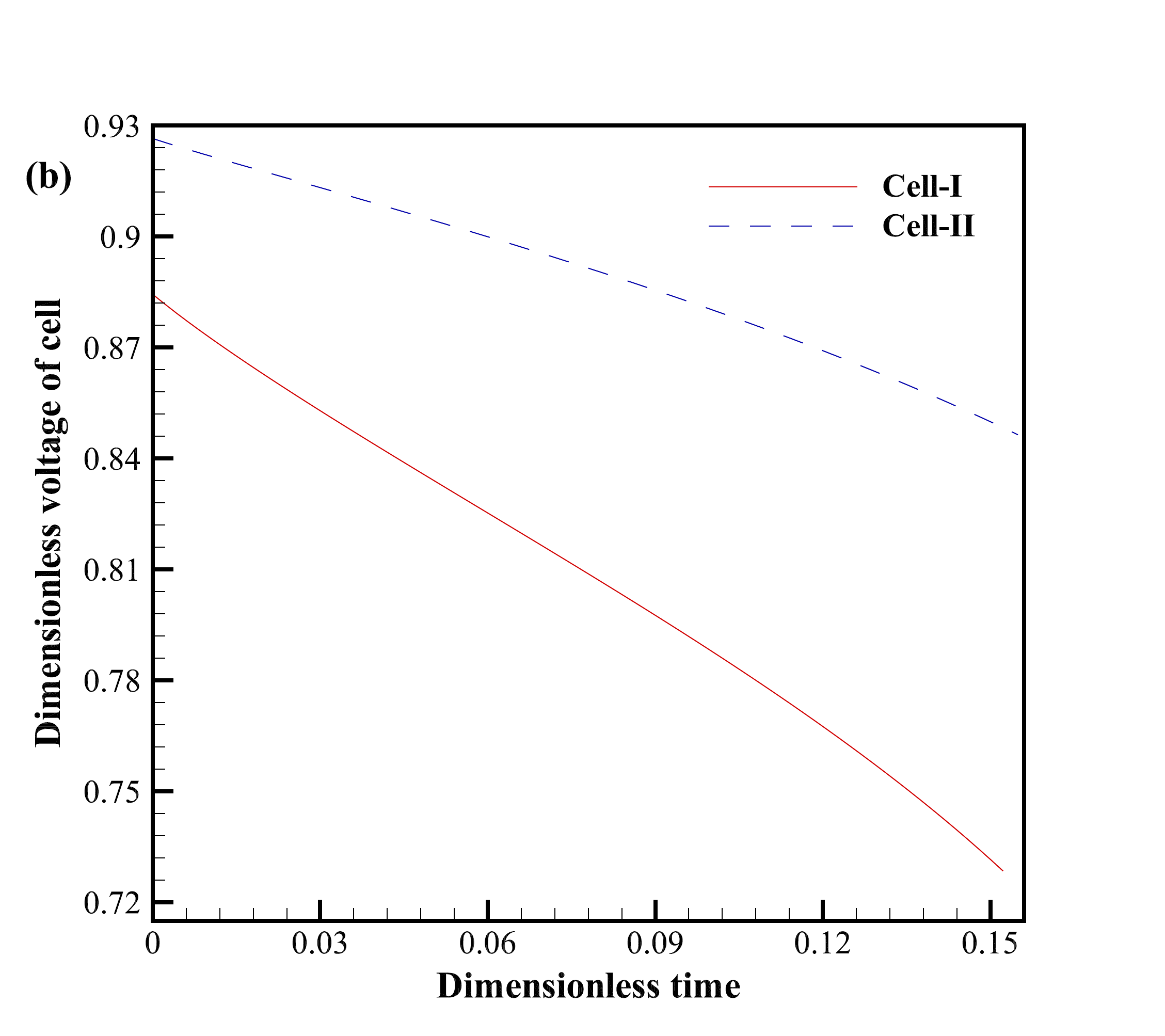}}
\caption{ Voltage of cell: (a) dimensional (b) dimensionless }
\label{vt}
\end{figure}

Discharging processes of two one--dimensional lead--acid cells have been simulated using finite volume method for both dimensional and non--dimensional systems of governing equations, in case study I. For case study II, discharging process have been conducted in three constant current rates, besides, the cell has been simulated using finite volume method too. In order to validate simulation results, voltage of cells has been compared with the same cell studied by Gu et al. \cite{guetal87} and Gu et al. \cite{guetal97} for case I, and with experimental data for case II. In \cref{validation} it can be seen a good consistency between the results which validate the numerical simulations. During discharge, the electric potential of the cell has been reached to cut--off voltage of 1.55 volt. Decreasing of cell voltage for  Cell--I and Cell--II over the time of discharge,  can be seen in \cref{vt}(a). The voltage  of Cell--I drops about two times faster than Cell--II because initial properties, operating conditions and geometry of the cells are totally different while their cut--off voltage is the same. ‌Beside, the discharge duration of Cell--I is shorter than Cell--II. Consequently, voltage of Cell--I decreases with faster slope comparing to Cell--II. In addition, Cell--I experienced wider range of voltage during shorter duration of time period. This figure give some useful information about the cells voltage but present no logical tool for comparison of each cell to it's maximum potential. 

In contrast, the results of non--dimensional cell voltage are plotted in \cref{vt}(b). As shown in the figure, dimensionless voltage of cell has been decreased for both batteries during discharge, while according to \cref{phisdless}, dimensionless voltage is the ratio of cell voltage to $V_{\mathrm{oc},0}$ of the battery. Then, the dimensionless voltage has been declined with respect to  $V_{\mathrm{oc},0}$. In this case, batteries have been evaluated in regard to the maximum voltage they could have. In fact, both batteries have been investigated relative to their maximum capabilities and this concept presents a more useful point of view for comparative purposes. In addition, dimensionless time for both batteries were almost equal accidentally and could be different in various batteries. This means that both batteries done their tasks in an equal relative time with respect to CTT. Therefore, higher dimensionless voltage amounts of Cell--II could demonstrate that Cell--II is more efficient than Cell--I concerning to $V_{\mathrm{oc},0}$. 

\begin{figure}
	\centerline{\includegraphics[trim=0 0 0 80mm ,width=0.5\textwidth]{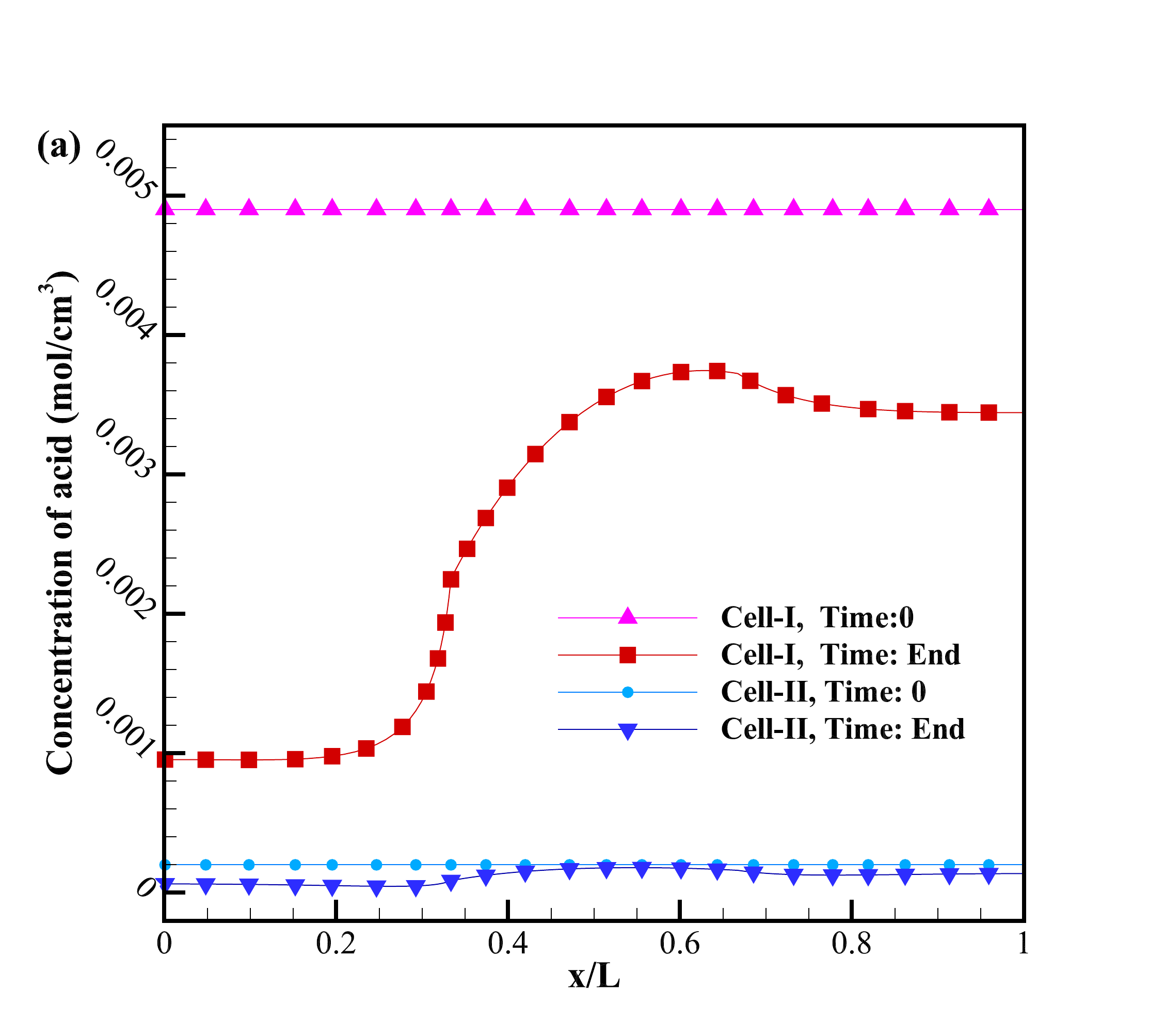}}
	\centerline{\includegraphics[trim=0 0 0 20mm, clip, width=0.5\textwidth]{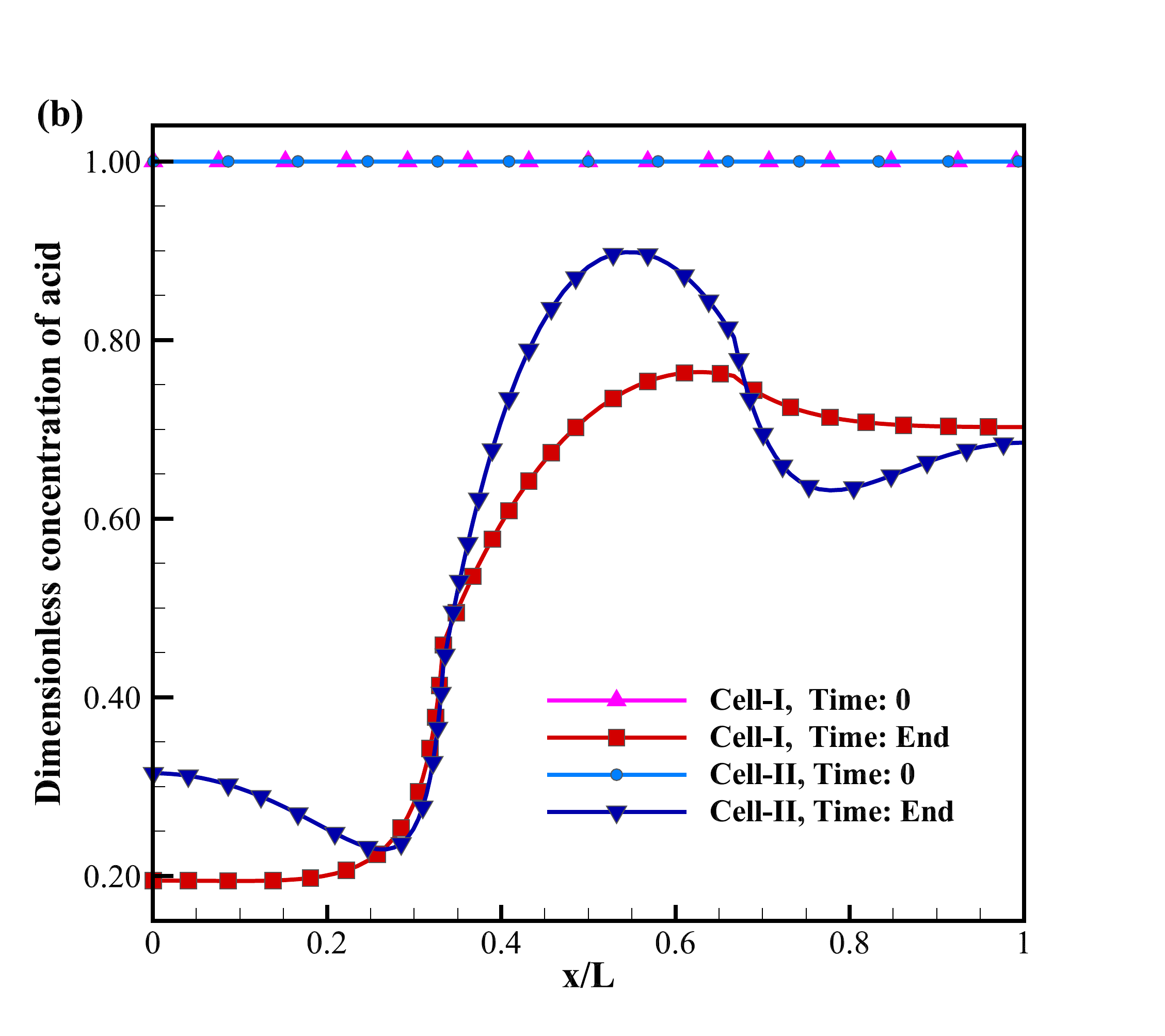}}
	\centerline{\includegraphics[width=0.5\textwidth]{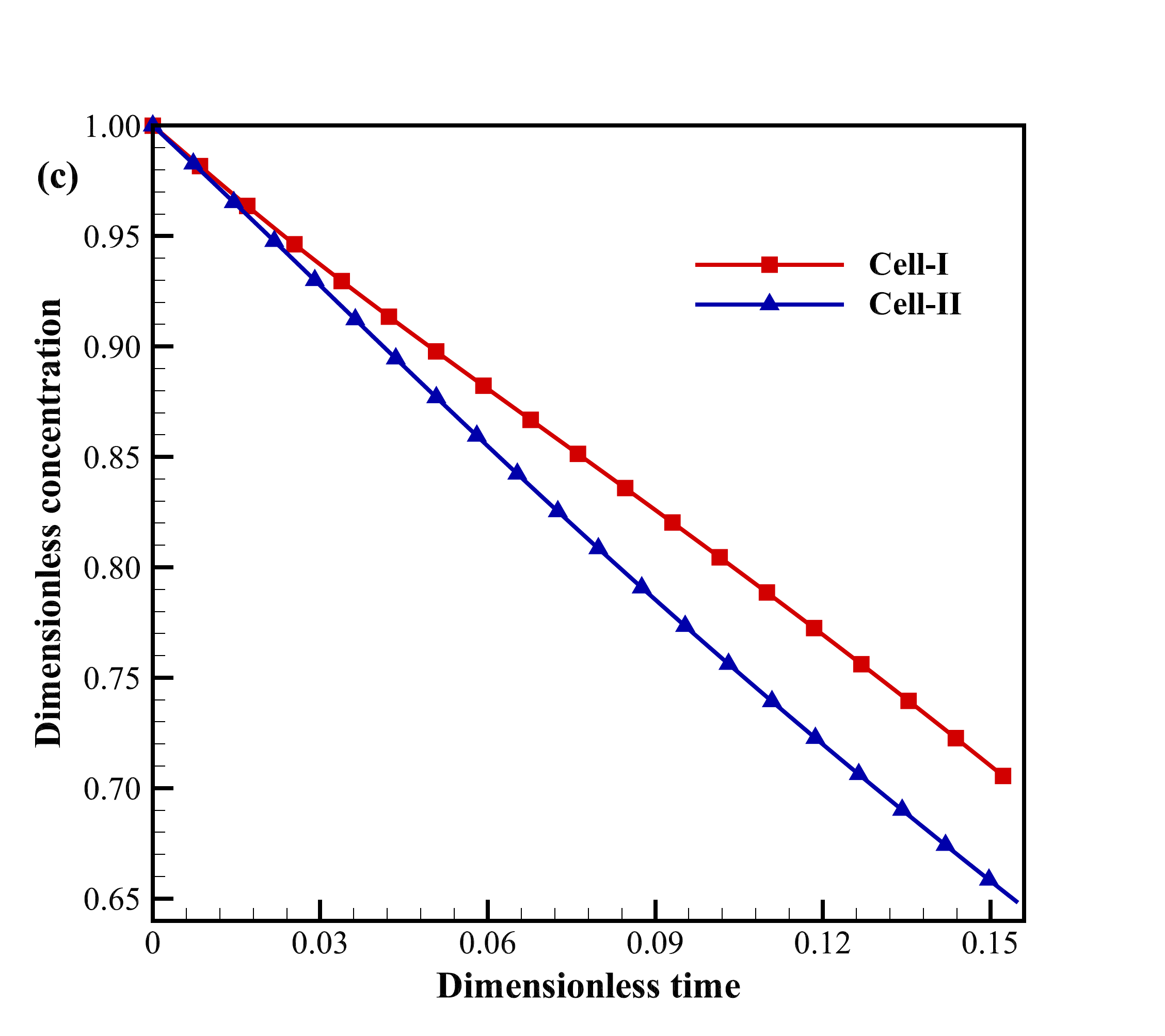}}
	\caption{Concentration of acid: (a)dimensional over cell (b)dimensionless over cell (c)dimensionless over time}
	\label{cx}
\end{figure}

In \cref{cx}(a) acid concentration along the cells are shown. As can be seen, plotted amounts for Cell--II have not good precision in comparison to Cell--I. Dimensionless concentration is presented in \cref{cx}(b). As can be seen in the figure, variation domain of $c^{\ast}$ is between 0 to 1 thus the results are conveniently comparable. Figure~(\ref{cx})(b) shows that non--dimensional concentration of Cell--I dropped more quickly than Cell--II in Region--I and Region--II during discharge. A cell can perform more desirable by decreasing concentration uniformly in the regions. Thus, comparison of concentration uniformity along each cell during discharge, can be a way to realize the better one. By calculating average amounts of $c^{\ast}$ of domain nodes, one can compare the average dimensionless concentration of Cell--I and Cell--II. Therefore, the calculation of average $c^{\ast}$ for both cells showed that acid concentration of Cell--I is about $\%6.2$ more uniform than the other one. In fact, this uniformity of plots is about the diffusion and it shows that the diffusion of Cell--I is better than Cell--II. 

As shown in \cref{cx}(c), $c^{\ast}$ decreases with dimensionless time during discharge. Decline trend of both cells are almost linear and slope of Cell--II is more than Cell--I that means in an equal time interval, concentration reduction of Cell--II is more than Cell--I. It is worth noting that this comparison is based on the workloads exerting on each cell and with changing load the results might be changed.

\begin{figure}
	\centerline{\includegraphics[trim=0 0 0 60mm, width=0.5\textwidth]{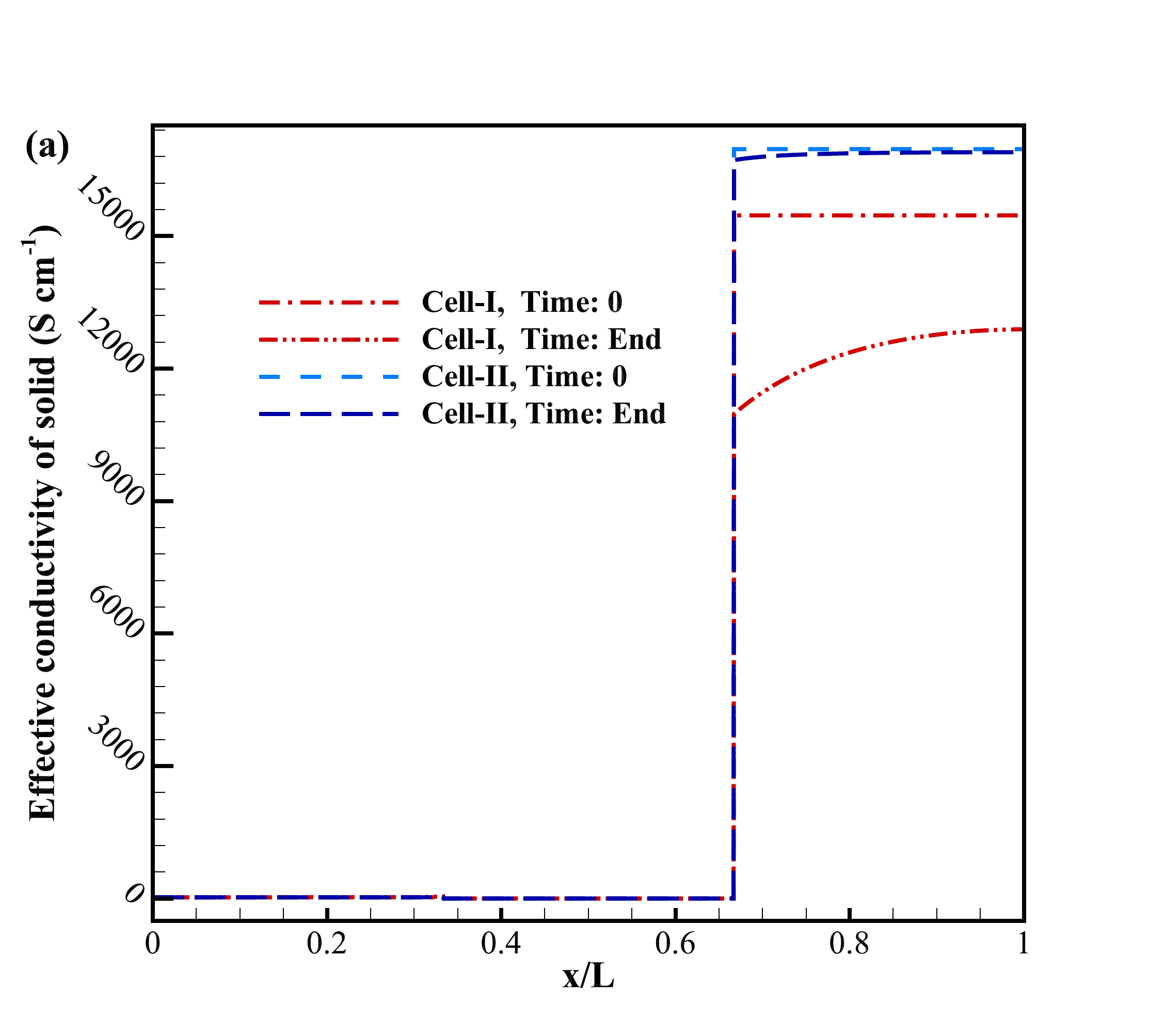}}
	\centerline{\includegraphics[trim=0 0 0 10mm,  clip, width=0.5\textwidth]{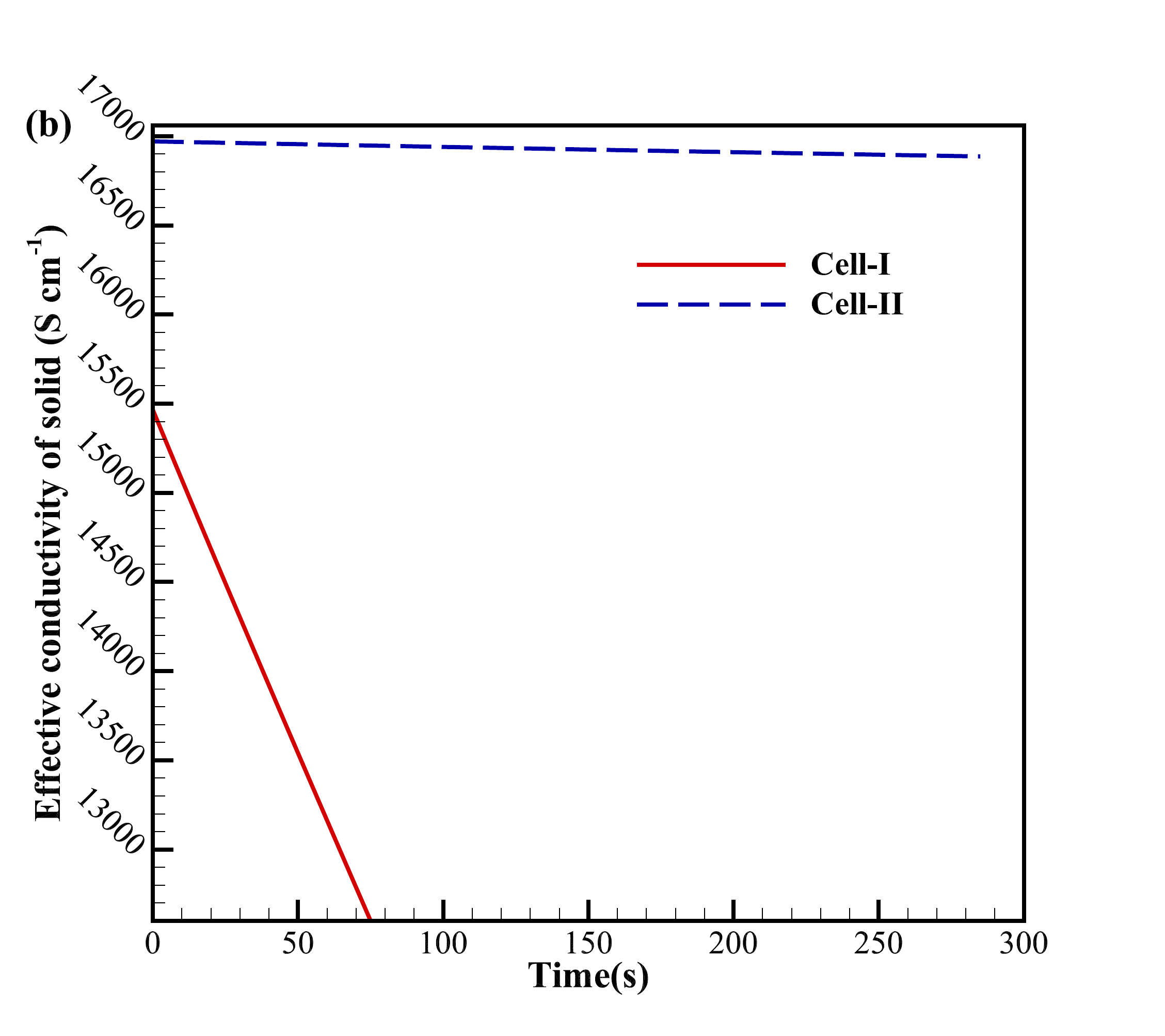}}
	\caption{Effective solid conductivity: (a)over length (b)over time}
	\label{sd}
\end{figure}

Figure~(\ref{sd})(a) illustrates that effective conductivity of solid decreases over the cells during discharge process. The decreasing of solid conductivity is because of porosity effect on effective conductivity due to composing lead sulphate. Variation of effective conductivity of solid versus time in the midpoint of Region--3 can be seen in \cref{sd}(b). The slope of both cells is linear and conductivity of Cell--I decreases faster. As can be seen in the figures, dimensional form can not present good comparison.

\begin{figure}
	\centerline{\includegraphics[trim=0 0 0 60mm, width=0.5\textwidth]{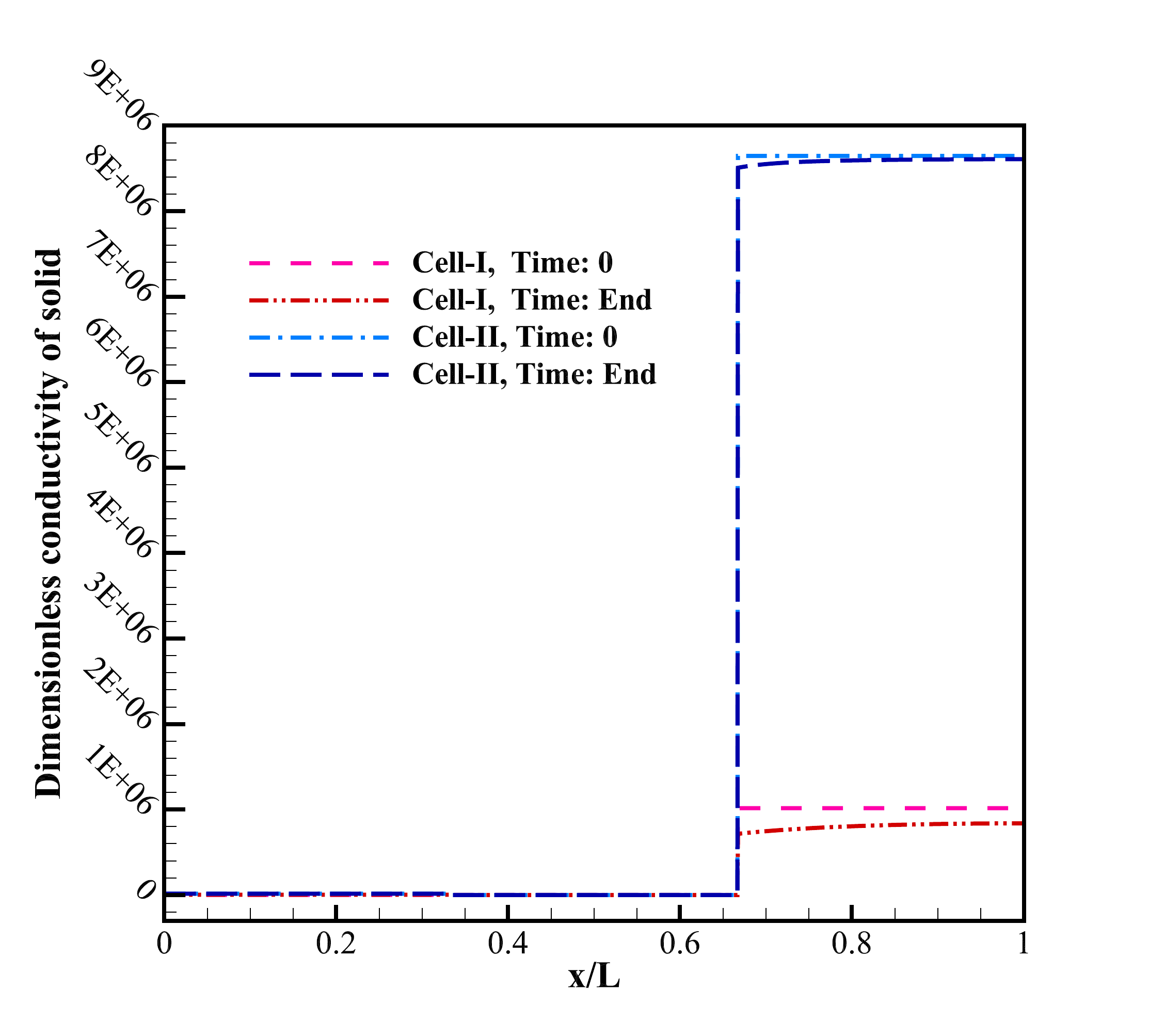}}
	\caption{Dimensionless conductivity of solid}
	\label{sxnd}
\end{figure}

As shown in \cref{sxnd}, $\sigma^{\ast}$ of Cell--II is in higher range than Cell--I and it is because of less amount of it's $\sigma_{\mathrm{exchange}}$. According to \cref{sigmaast1,sigmaast2,sigmaast3}, $\sigma_{\mathrm{exchange}}$ is a defined conductivity while exerting $V_{\mathrm{oc,0}}$ with current of $i_0 A_{\mathrm{max} }L$. The main reason for less amount of $\sigma_{\mathrm{exchange}}$ in Cell--II is smaller exchange current density. In an other point of view, the higher values of $\sigma^{\ast}$ means higher value of $\sigma^{\mathrm{eff}}$. In result, the conductivity of Cell--II is more dependent on the initial conductivity and the cell is in more active state.

\begin{figure}
	\centerline{\includegraphics[trim=0 0 0 60mm, width=0.5\textwidth]{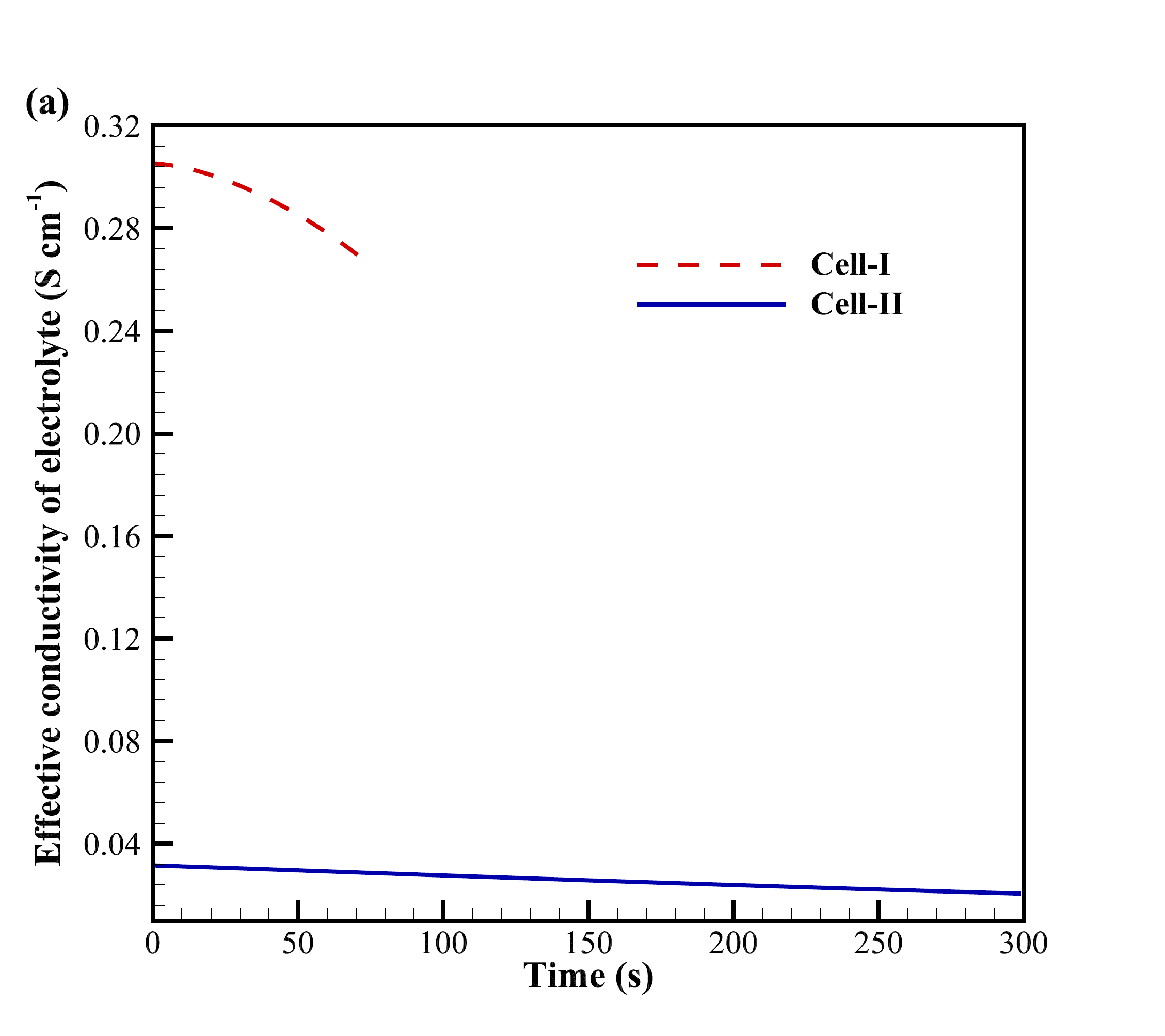}}
	\centerline{\includegraphics[trim=0 0 0 19mm, clip, width=0.5\textwidth]{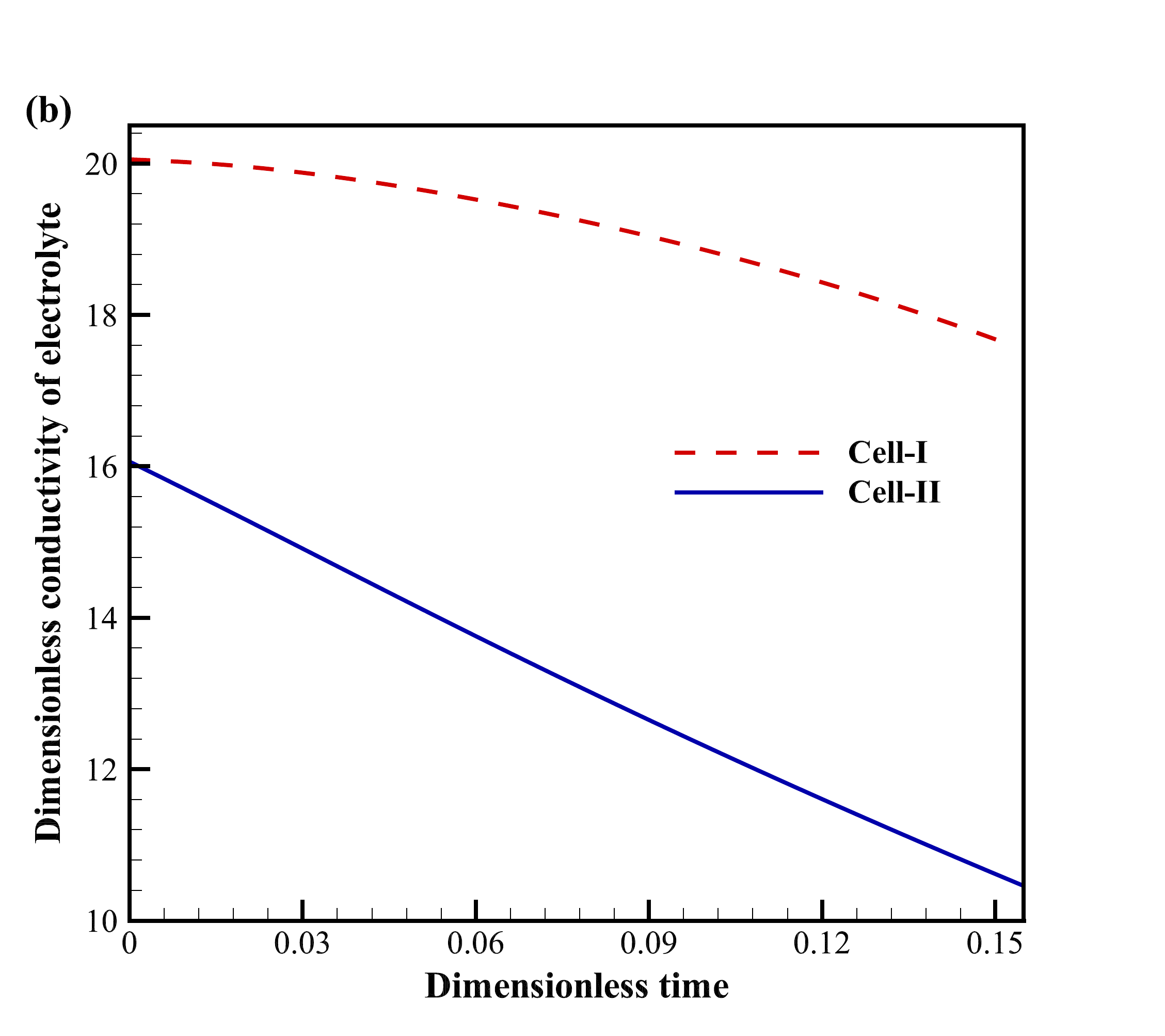}}
	\caption{Effective conductivity of electrolyte: (a)dimensional (b)dimensionless}
	\label{kd}
\end{figure}

Effective conductivity of electrolyte for both cells are shown in \cref{kd}(a). Amounts of effective conductivity for Cell--I are about ten times more than Cell--II. As can be seen, $k^{\mathrm{eff}}$ decreases during discharge and amounts of $k^{\mathrm{eff}}$ for Cell--I declines in wider range and shorter time interval than Cell--II. The slope of $k^{\mathrm{eff}}$ plot for Cell--II is almost linear while for Cell--I is non--linear and its slope increased gently to the end of process. All these explained physical phenomena are depended on variations of concentration and porosity according to \cref{keff}. The effective conductivity of electrolyte under constant temperature assumption is calculated as follows:

\begin{equation}
k^{\mathrm{eff}}= c \exp{}    \Biggl\{  \left( 22.3684 \right) + \left( 532.8843 \right) c - 16097/781 c^2   \Biggr\}
 \varepsilon^{\mathrm{ex}}
\label{keff}
\end{equation}
in which $\varepsilon$ is porosity and $\mathrm{ex}$ is a constant power. 

As can be seen in \cref{kd}(b) non--dimensional conductivity of Cell--I is about one and half times more than Cell--II. 
According to \cref{kast3}, the higher amounts of $k^{\ast}$ means the higher amounts of $k^{\mathrm{eff}}$ over $k_{\mathrm{exchange}}$. This fact shows that Cell--I is more active than Cell--II in the case of conductivity. Non--dimensional conductivity of Cell--I has experienced values of 20 to 18, from OCFCS to discharged state, respectively, and Cell--II has varied from 16 to about 10. That means $k^{\mathrm{eff}}$ of Cell--I is 20 to 18 times of $k_{\mathrm{exchange}}$ and for Cell--II $k^{\mathrm{eff}}$ is about 16 to 10 times of its $k_{\mathrm{exchange}}$. In another point of view,  \cref{kast1} shows that $k^{\ast}$ is the magnitude of $i_{\mathrm{oc,0}}$ into $i_{\mathrm{exchange}}$. It is important to note that $i_{\mathrm{oc,0}}$ is a assumptive current can be created by exerting $V_{\mathrm{oc,0}}$ and with conductivity of $k^{\mathrm{eff}}$. Wider range of $k^{\ast}$ for Cell--II illustrates that Cell--II loses its conductivity faster than Cell--I during discharge.

\begin{figure}
	\centerline{\includegraphics[trim=0 0 0 60mm, width=0.5\textwidth]{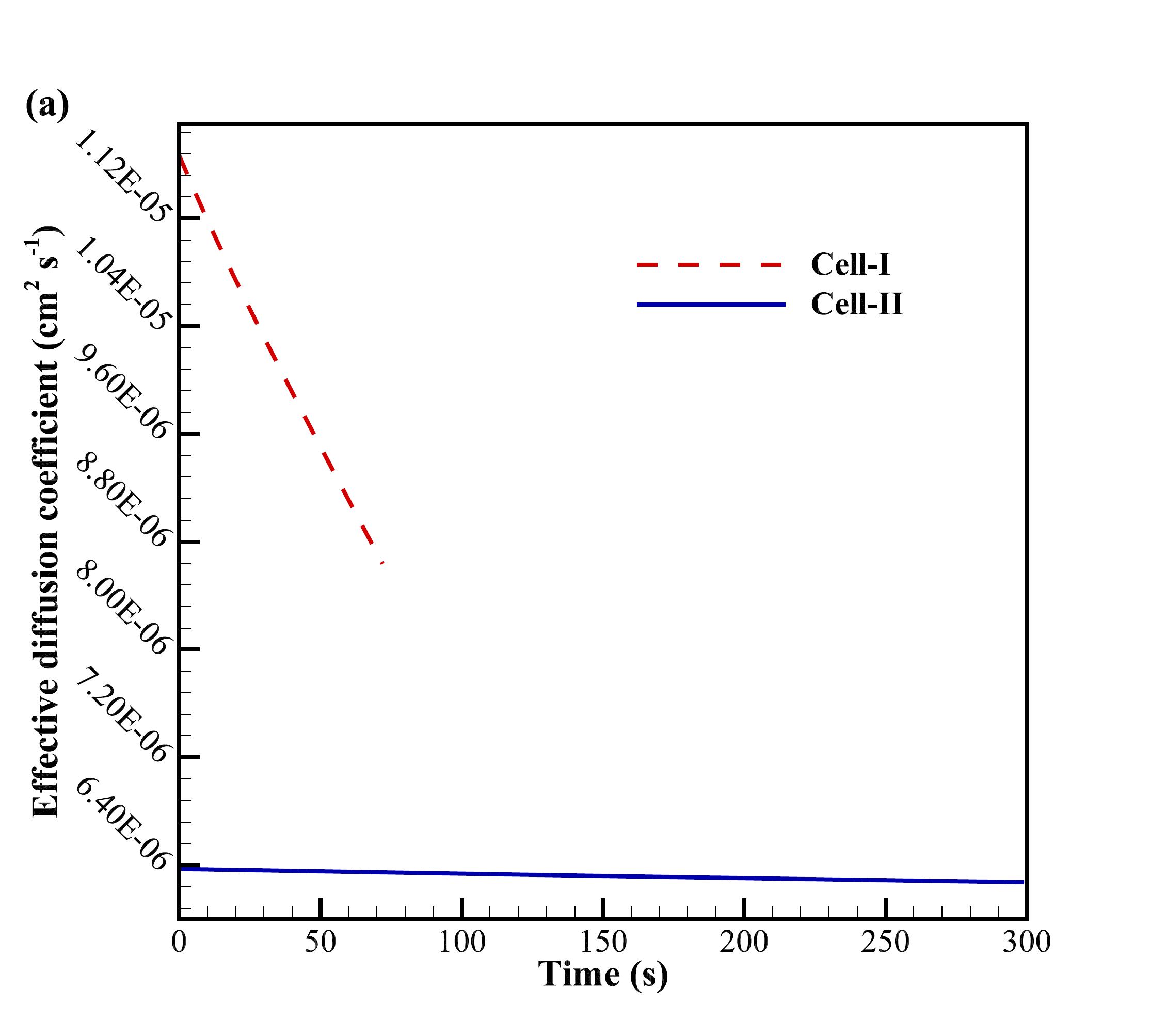}}
	\centerline{\includegraphics[trim=0 0 0 20mm, clip, width=0.5\textwidth]{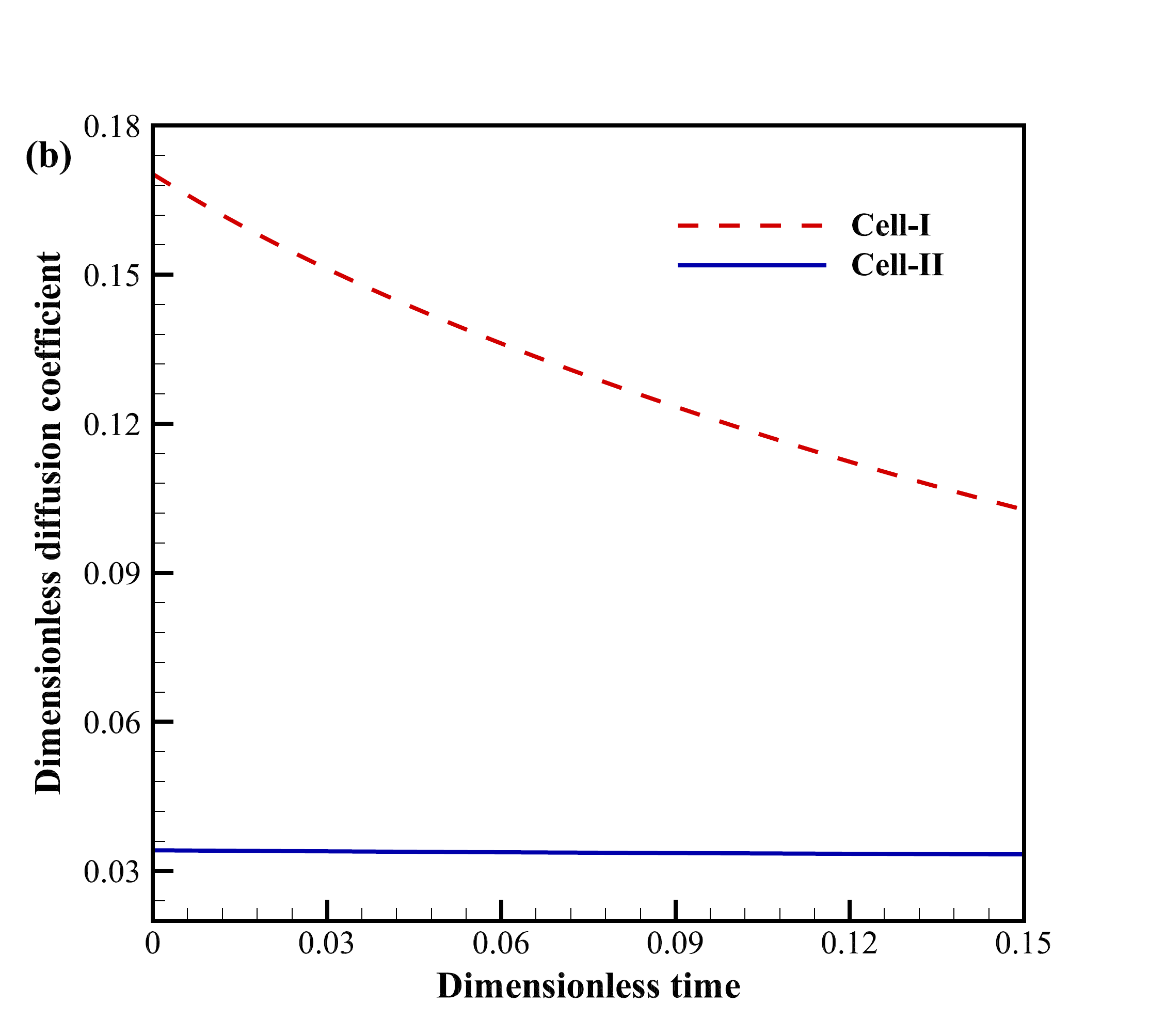}}
	\caption{Effective diffusion coefficient: (a)dimensional (b)dimensionless}
	\label{Dd}
\end{figure}

In \cref{Dd}(a) diffusion coefficients of the cells are presented. As can be seen in the figure, coefficients of Cell--I has declined rapidly and covered wider range of diffusivity. In contrary, diffusion coefficients of Cell--II decreases very slowly. However, diffusion coefficients of both cells are in order of $10^{-6}$ and very close to each other.

As can be seen in \cref{Dd}(b) non--dimensional diffusivity of cells take some distance from zero and become more comparable. Nevertheless, main reason of $D^{\ast}$ definition is to compare each cell to its OCFCS. According to \cref{dast1} effective diffusion of Cell--I has declined relative to $D_{\mathrm{exchange}}$ during discharge while for Cell--II remained almost constant. From \cref{dast2} there is another useful point of view that is $c_{\mathrm{exchange}}$ of Cell--I has increased during discharge while for Cell--II, $c_{\mathrm{exchange}}$ remained almost equal to $c_0$. From analyses of diffusivity of cells can be resulted Cell--I is more active and Cell--II is  more stable.

\begin{figure}
	\begin{subfigure}{0.5\textwidth}\centerline{\includegraphics[trim= 0mm 0mm 0mm 70mm, width=1.1\textwidth]{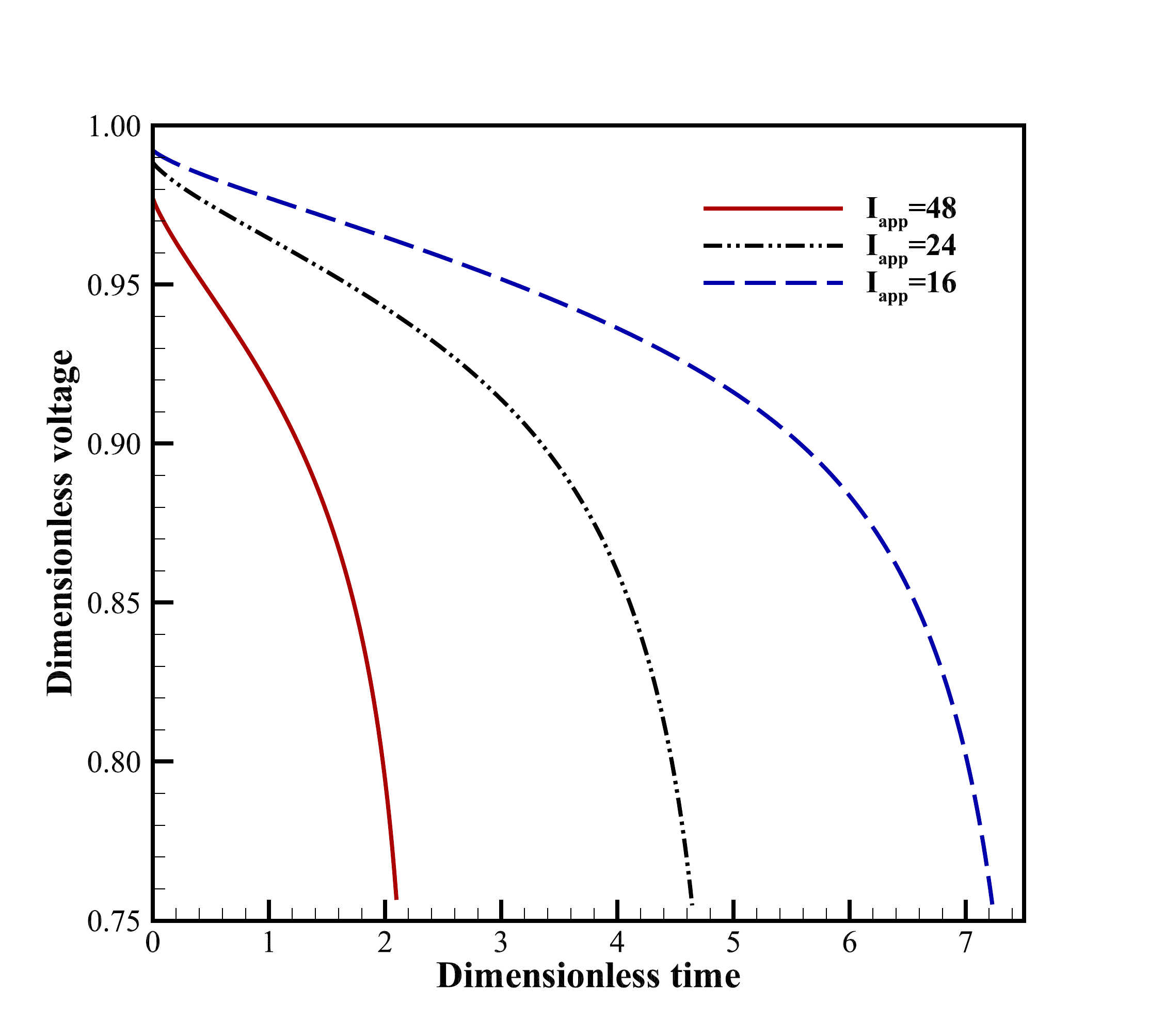}}
		\caption{dimensionless voltage}
		\label{Exp:v}
	\end{subfigure}
	\begin{subfigure}{0.5\textwidth}\centerline{\includegraphics[trim= 0mm 0mm 0mm 70mm, width=1.1\textwidth]{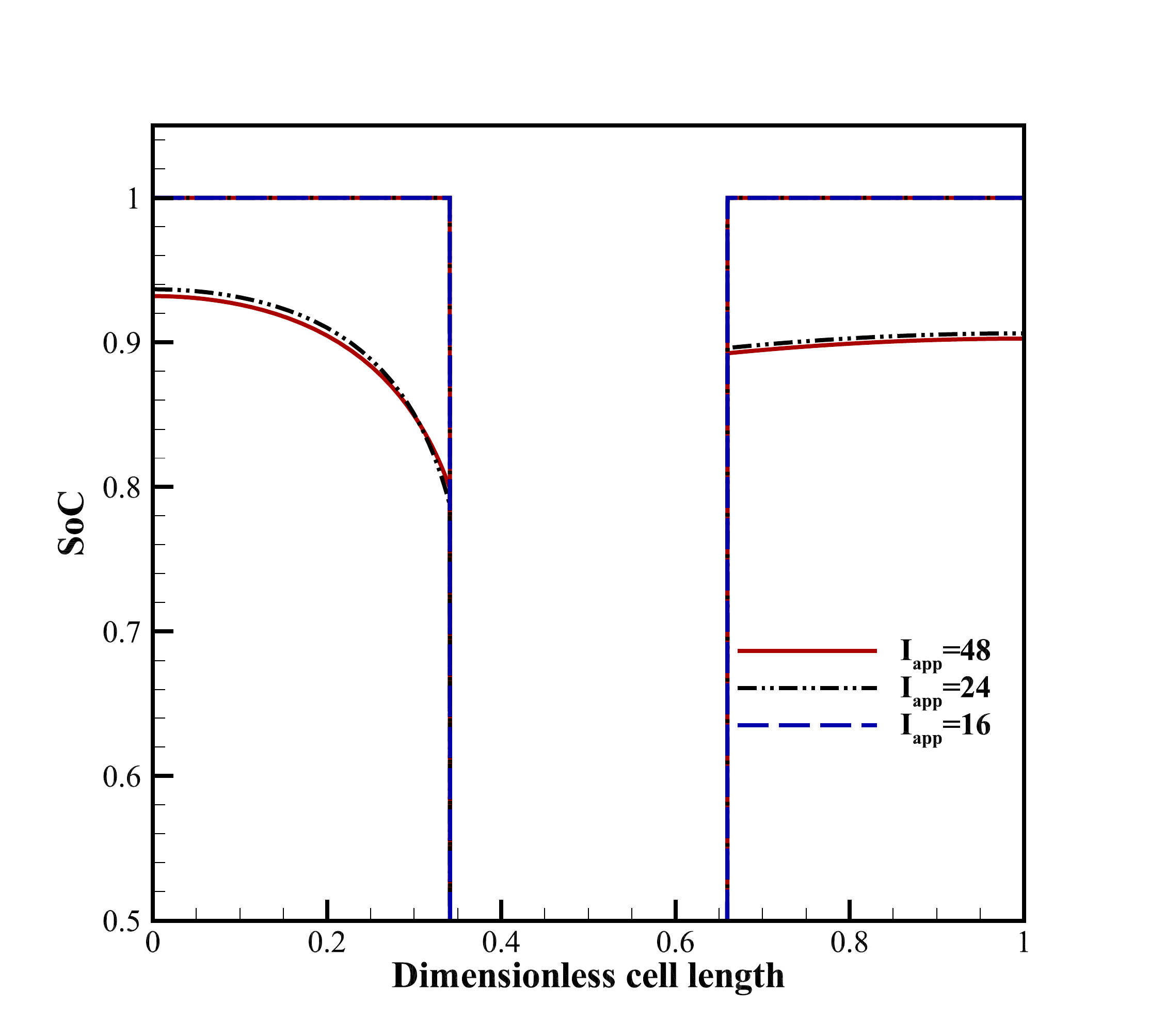}}
		\caption{state of charge}
		\label{Exp:soc}
	\end{subfigure}
	\begin{subfigure}{.5\textwidth}\centerline{\includegraphics[trim= 0mm 0mm 0mm 10mm, clip, width=1.1\textwidth]{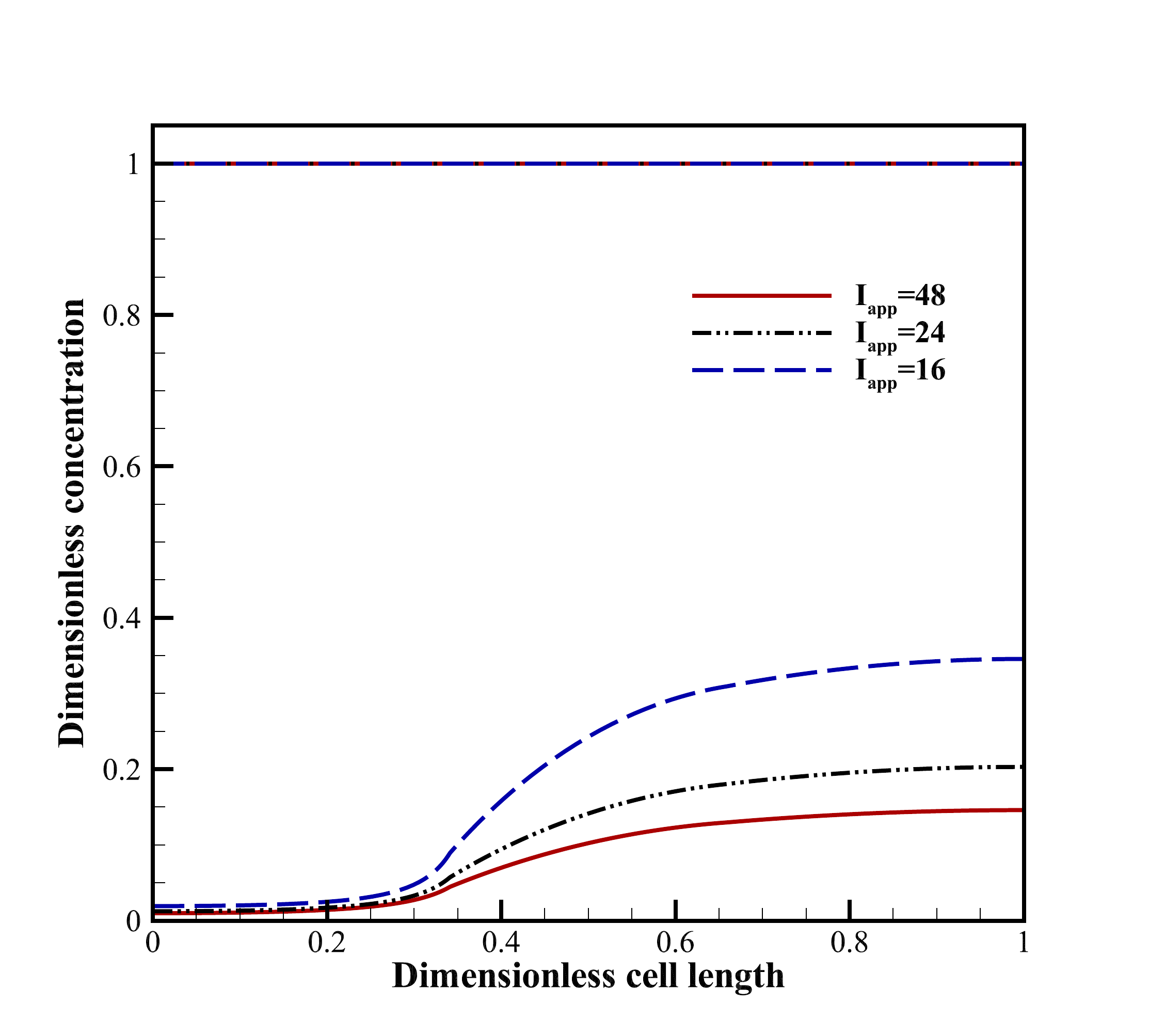}}
		\caption{dimensionless concentration}
		\label{Exp:c}
	\end{subfigure}
	\begin{subfigure}{.5\textwidth}\centerline{\includegraphics[trim= 0mm 0mm 0mm 5mm, width=1.1\textwidth]{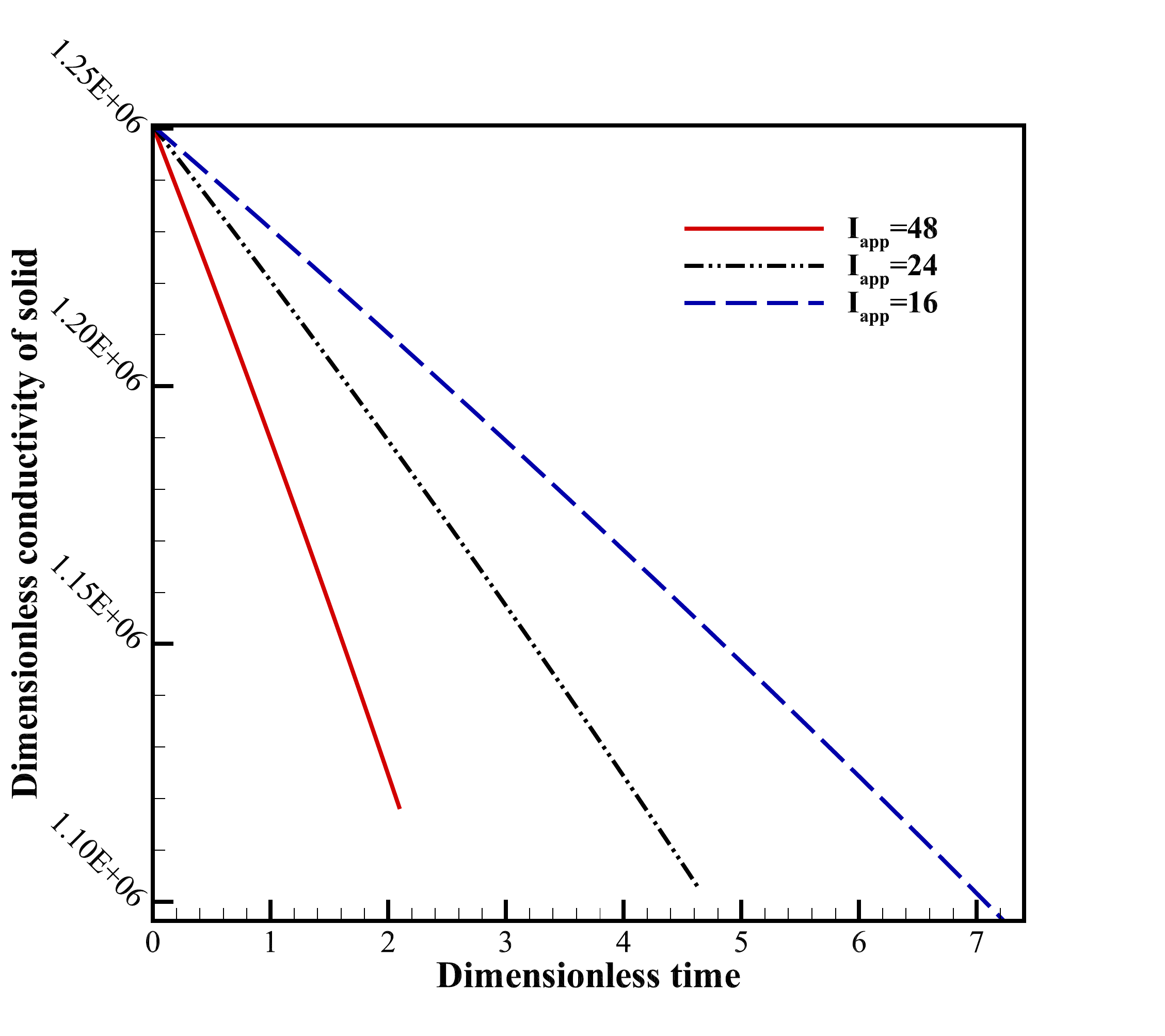}}
		\caption{dimensionless conductivity of solid}
		\label{Exp:sigma}
	\end{subfigure}
	\begin{subfigure}{.5\textwidth}\centerline{\includegraphics[trim= 0mm 0mm 0mm 10mm, clip, width=1.1\textwidth]{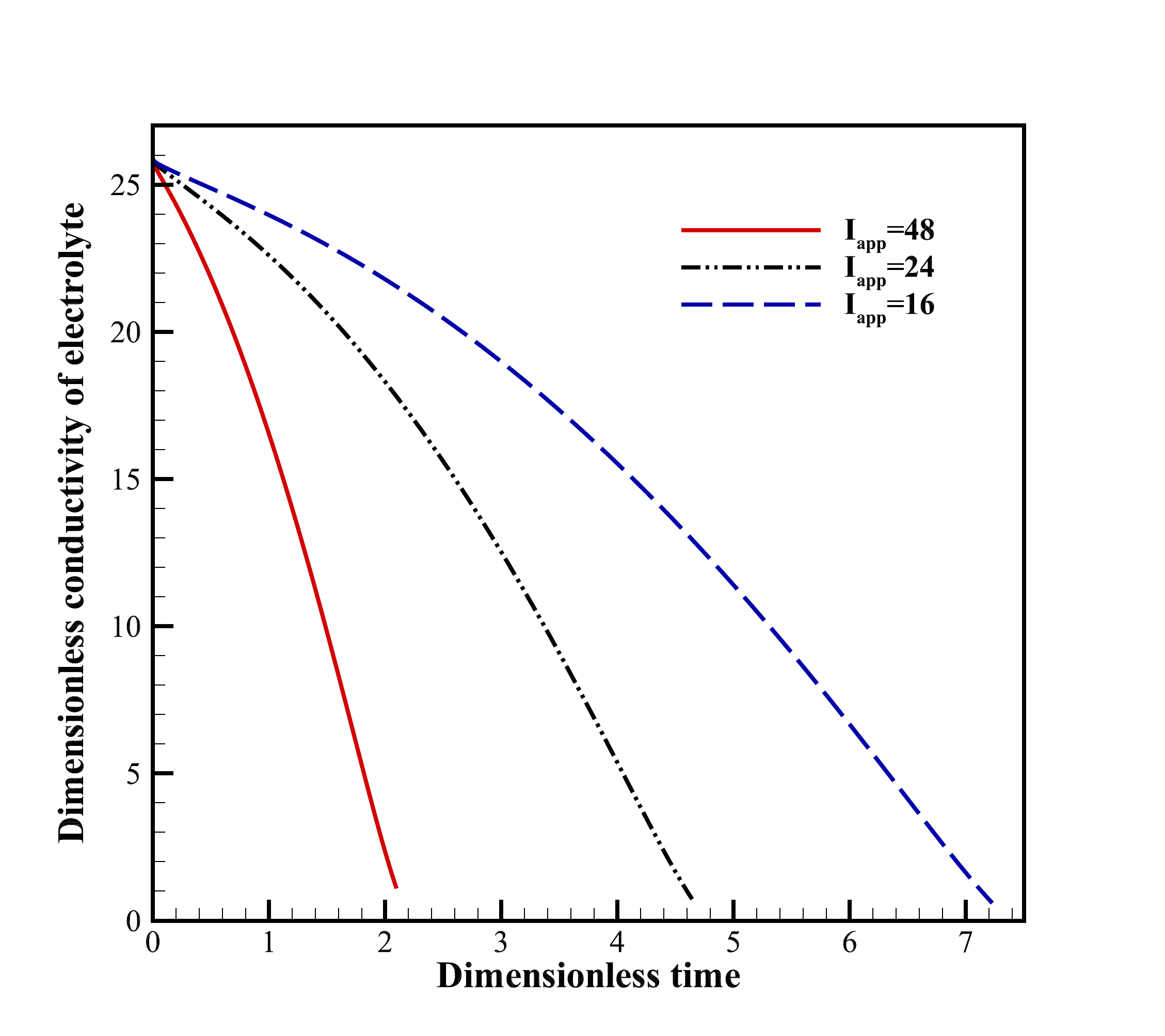}}
		\caption{dimensionless conductivity of electrolyte}
		\label{Exp:k}
	\end{subfigure}	
	\begin{subfigure}{.5\textwidth}\centerline{\includegraphics[trim= 0mm 0mm 0mm 6mm, width=1.1\textwidth]{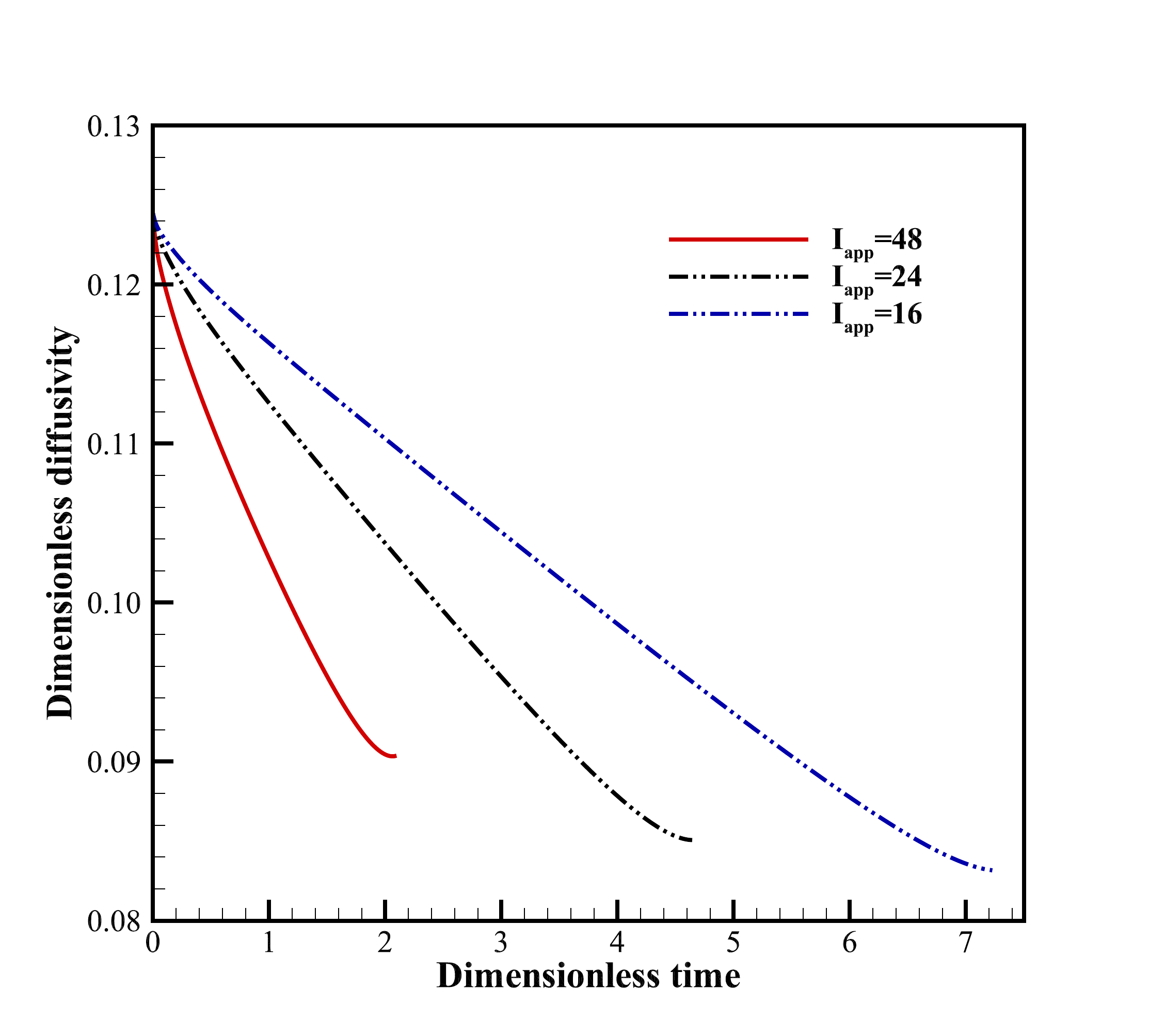}}
		\caption{dimensionless diffusivity}
		\label{Exp:D}
	\end{subfigure}	
	\caption{ Variations of some properties versus dimensionless time and cell--length in three levels of constant current density ($16$ mA cm$^{-2}$, $24$ mA cm$^{-2}$ and $48$ mA cm$^{-2}$) during discharge. }
	\label{Exp}
\end{figure}

The results of case study II, are shown in \cref{Exp}.
Moreover, the results of experiment data are shown in \cref{validation} to valid the simulation tests.
 As can be seen in \cref{Exp:v} dimensionless voltage declines during dimensionless time in three levels of applied current density: $16$ mA cm$^{-2}$, $24$ mA cm$^{-2}$ and $48$ mA cm$^{-2}$. In addition, by increasing applied current density, $V^{\ast}$ and $t^{\ast}$ decreases. The figure shows that all plots, in an efficient behavior, started from a point above 97\% of $V_{\mathrm{oc},0}$ and at the end reached to 75\% of their initial potential. Moreover, time durations of discharging processes are approximately 2, 4.5 and 7 times more than CTT for 16, 24 and 48 amperes per cubic centimeter, respectively. \Cref{Exp:soc} illustrates state of charge in both electrodes and as can be seen it is almost equal in every points of the domain. Dimensionless concentration of electrolyte falls down during discharge as can be seen in \cref{Exp:c}. In Region--1 the acid is almost consumed, but in Region--3 and in lower values of discharging rate, the higher amounts of acid remains not consumed. For applied current density of 48mA$/$cm$^2$, the remained acid is more than 10\% of initial concentration, while in two other applied currents the amount of concentrations are about 20\%  and 30\% of initial concentration, respectively. 

In \cref{Exp:sigma}, dimensionless conductivity of solid increases with increasing time, similarly, it takes higher values in higher amounts of applied current density. Conversely, $i_{\mathrm{oc}(s)}$ have higher amounts proportional to $i_{\mathrm{exchange}}$ in lower applied currents at the end of process. \Cref{Exp:k} shows variations of dimensionless conductivity of electrolyte to dimensionless time. Over time, conductive current density of electrolyte falls down related to ECD. Furthermore, by reducing the applied current density, the conductive current density increases proportional to ECD.
\Cref{Exp:D} represents dimensionless diffusivity decreasing versus dimensionless time. As can be seen in the figure, diffusional molar flow rate reduces than molar flow rate of OCFCS. The minimum amount of diffusional molar flow rate reduces than molar flow rate of OCFCS in the end of discharging process and in minimum $I_{\mathrm{app}}$. However, for every discharge rate it declines under 0.09. 

\section{Conclusion}
Equations set of charge conservation in electrode and electrolyte and conservation of species for lead--acid batteries are non--dimensionalized by determination of some dimensionless parameters. Dimensionless coefficients of $\sigma^{\ast}$, $k^{\ast}$, $k_D^{\ast}$ and $D^{\ast}$ are resulted from non--dimensionalization process. The open circuit fully charged state (OCFCS) is assumed to be a base state for definition of the dimensionless coefficients. 
The main reason for determination of OCFCS is each battery should be evaluated with its maximum potential and the results of this evaluation could be compared between batteries. 
According to results, dimensionless voltage and solid conductivity of Cell--II was better than Cell-I, while in the cases of dimensionless acid concentration, electrolyte conductivity and diffusional coefficient, Cell-I was the better one. In conclusion, Cell--I is preferable despite shorter time duration of discharge and lower amounts of solid dimensionless conductivity. In addition, from \cref{sigmaast}, $\sigma^{\ast}$ could have higher values by increasing $V_{\mathrm{oc},0}$ and decreasing $i_0$, $A_{\mathrm{max}}$ and $L$. As can be concluded from \cref{sigmaast}, the cell length with power of two is the most effective parameter. Thus, by changing geometry and structure of a cell one can improve dimensionless solid conductivity as well as  coefficients of $k^{\ast}$ and $D^{\ast}$ according to \cref{kast} and \cref{dast}. 
Furthermore, the results demonstrated that dimensionless analyze and using dimensionless coefficients facilitates elechtrochemical analyses and give more useful concept of physical problem. Finally, it is worth noting that the non--dimensional model have benefits of dimensional model and moreover, gives new concepts of battery behavior plus some novel points of view.

\section*{References}

\bibliography{mybibfile}

\section*{Glossary}
$A$: specific electroactive area (cm$^2$ cm$^{-3}$)

$c$: acid concentration (mol cm$^{-3}$)

$c_0$: initial acid concentration (mol cm$^{-3}$)

$D$: diffusion coefficient (cm$^2$ s$^{-1}$)

$F$: Faraday constant, 96487 C mol$^{-1}$

$i_0$: exchange current density (A cm$^{-2}$)

$j$: transfer current density (A cm$^{-2}$)

$k$: conductivity of liquid (S cm$^{-1}$)

$L$: cell length (cm)

$R$: universal gas constant, 8.3143 J mol$^{-1}$ k$^{-1}$

$t$: time (s)

$T$: temperature (K)

$V$: cell voltage (V)

Greek letters

$\alpha_a, \alpha_c$: anodic and cathodic transfer coefficient

$\varepsilon$: porosity

$\eta$: electrode overpotential (V)

$\sigma$: conductivity of solid (S cm$^{-1}$)

$\phi$: electric potential (V) 

Subscripts and super scripts

$D$: pertinent to diffusion

$\mathrm{eff}$: effective

$e$: electrolyte

$\mathrm{max}$: maximum

$oc,0$: open circuit at time zero

$\mathrm{ref}$: reference

$s$: solid

\end{document}